**A Comprehensive Review of Deep Learning Applications in Hydrology and Water Resources**

Muhammed Sit, Bekir Z. Demiray, Zhongrun Xiang, Gregory J. Ewing, Yusuf Sermet, Ibrahim Demir

**Abstract**

The global volume of digital data is expected to reach 175 zettabytes by 2025. The volume, variety, and velocity of water-related data are increasing due to large-scale sensor networks and increased attention to topics such as disaster response, water resources management, and climate change. Combined with the growing availability of computational resources and popularity of deep learning, these data are transformed into actionable and practical knowledge, revolutionizing the water industry. In this article, a systematic review of literature is conducted to identify existing research which incorporates deep learning methods in the water sector, with regard to monitoring, management, governance and communication of water resources. The study provides a comprehensive review of state-of-the-art deep learning approaches used in the water industry for generation, prediction, enhancement, and classification tasks, and serves as a guide for how to utilize available deep learning methods for future water resources challenges. Key issues and challenges in the application of these techniques in the water domain are discussed, including the ethics of these technologies for decision-making in water resources management and governance. Finally, we provide recommendations and future directions for the application of deep learning models in hydrology and water resources.

## Introduction

The global volume of digital data is expected to reach 175 zettabytes by 2025 (Reinsel et al., 2018). Large-scale sensor networks as well as the increased awareness of climate change, water resources management, and the monitoring of water-related hazards led to the substantial growth of the volume, variety, and velocity of water-related data (Weber et al., 2018; Sit et al., 2019). Modern data collection techniques, including satellite hydrology, internet of things for on-site measurements (Kruger et al., 2016), and crowdsourcing tools (Sermet et al., 2020b), has revolutionized the water science and industry as approached by the government, academia, and private sector (Krajewski et al., 2016). The effective utilization of vast water data holds the key for long-term sustainability and resilience and presents opportunities to transform water governance for the upcoming decades (Grossman et al., 2015). In the hydrological domain, multivariate analysis relying on extensive and semantically-connected data resources is required to generate actionable knowledge and produce realistic and beneficial solutions to water challenges facing communities (Jadidoleslam et al., 2019; Carson et al., 2018). However, the inaccessible, unstructured, nonstandardized, and incompatible nature of the data makes optimized data models (Demir and Szczepanek, 2017) and smarter analytics approaches a necessity (Sermet and Demir, 2018a).

Computerized methods to create an understanding of hydrological phenomena are based on various modeling strategies, which simplify a hydrological system to simulate its behavior (Antonetti and Zappa, 2018). Physical models aim to achieve this goal by specifically designing complex simulations that are powered by mathematical and numeric specifications of conceptualized physical characteristics (Jaiswal et al., 2020). However, hydrological systems, as is the case with other natural systems, are inherently heterogeneous (Marçais and De Dreuzy, 2017) as opposed to less complex human-made systems with defined rules. Therefore physical models, though are deterministic and reliable, do not always perform and scale well due to their intrinsic limitations (Islam, 2011). As an alternative, statistical models have been employed to make use of the comprehensive set of available hydrological, environmental, and geophysical data (Evora and Coulibaly, 2009). These approaches assume minimum awareness of the underlying mechanism and receive their strength by eliciting useful information and patterns from the available data through statistical analysis (McCuen, 2016). However, they have displayed shortcomings in terms of accuracy and certainty, and also require excessive computational power (Ardabili et al., 2019; Agliamzanov et al., 2020).

Recent developments in artificial intelligence and graphical processor units (GPU) have paved the way for deep learning, a pioneering approach that is fueled by multilayer artificial neural networks (LeCun et al., 2015). Deep learning provides a black-box method to learn from complex and high-dimensional data to infer robust and scalable insights while minimizing the degree to which manual labor is needed (Sengupta et al., 2020). One feature which separates deep learning from its superset machine learning is the use of multilayer models which leads to a higher-level representation of the underlying data sources (Saba et al., 2019). Furthermore, deep learning is capable of extracting substantial features without being explicitly instructed, and thus, is more immune to raw and noisy data (Sahiner et al., 2019). Successful implementations of deep learning permeate numerous domains and industries including medical imaging (Haskins et al., 2020), healthcare (Esteva et al., 2019), finance (Heaton et al., 2017), geophysical sciences (Shen, 2018), remote sensing (Ma et al., 2019), and hydrology. Due to its significant adoption rate and potential to be applicable to any domain which encompasses problems that can be expressed as control systems, numerous open-source and for-profit software tools, educational resources, and generalized algorithms have been made available for use, opening up countless paths to advance hydrological studies.

This paper presents a systematic review of applications of deep learning within the scope of the hydrological domain. The literature has been thoroughly examined to identify the use cases of deep learning in the subfields of the water sector including flooding, weather, land use and soil, water quality, surface water, water resources, and groundwater. Each study has been evaluated to extract information that is scientifically relevant to assess the study's contribution and reproducibility including the hydrological tasks that were taken on to be approached by deep learning along with the utilized network architectures, datasets, software tools and frameworks, licenses, and deep learning practices and algorithms. The paper explores modern deep learning networks from the lens of the hydrological domain to investigate the shortcomings and challenges of the status quo and to highlight the opportunities and future directions to serve as a guide to researchers, professionals, and governmental organizations in the water sector.

The major contributions of this paper can be summarized as follows. Though there are various configurations of artificial neural networks optimized for various data types and use cases, it is challenging to reduce a real-life hydrological task to a certain predefined approach given the depth and complexity of the tasks as well as the diversity of networks. The methodologies that need to be employed while developing deep learning-powered solutions in hydrology are not standardized in terms of data quality and preparation, execution, validation, and documentation. Furthermore, the strength, usability, and reliability of a model lie on clearly set descriptions and procedures for deterministic reproducibility, given the variety of development frameworks as well as the application areas. To the best of our knowledge, there has not been a thorough investigation of systematically approaching water challenges with deep learning. Thus, this paper serves as a meticulous guide for the stakeholders of the hydrology domain to advance the water industry with intelligent systems that are revolutionized by multi-faceted data (Sermet et al., 2020a).

The remainder of this article is organized as follows. *Literature Review* section provides the review methodology followed by a comprehensive literature review of deep learning applications in the water domain. Descriptions of deep learning concepts, tasks, and architectures are described to summarize the available methodology for use by the hydrological community. *Results* section presents a detailed summary and analysis of reviewed papers grouped by their application area. *Key Issues and Challenges* section highlights the key issues and challenges facing the stakeholders utilizing deep learning in the water domain with respect to technical limitations as well as ethical considerations. *Recommendations and Conclusions* section outlines a vision entailing the adoption of prominent and deep learning-powered technologies to solve the water challenges of the future and then, concludes the paper with a concise summary of findings.

## Literature Review

This section starts with a detailed description of the literature search methodology in the first subsection *Review Methodology* and then presents the information extracted from each reviewed manuscript. The subsection *Deep Learning* gives a brief overview of deep learning history, describes various neural network architectures, and elaborates on different machine learning task types. At the end of this section we share the summary of the literature as figures to provide an understanding of this review and a table of all the papers reviewed.

### Review Methodology

A systematic literature search on water domain was employed for this review. Web of Science, Scopus, Springer Link, Wiley Online Library and The International Water Association Publishing Online were used as the databases and the keywords included "deep neural network", "deep neural networks", "deep learning", "lstm", "long short term memory", "cnn", "convolutional", "gan", "generative adversarial", "rnn", "recurrent neural", "gru" and "gated recurrent". After limiting the search with these keywords in publication title, abstract or keywords, an additional exclusion criterion was applied through each database's categorization system if applicable only to include the publications within the environmental fields. Also, searches were limited to only include journal publications. All articles published in 2018, 2019 and 2020 up until the end of March containing these keywords in their titles, keyword fields or abstracts were included in the first list of articles gathered. This time interval is primarily chosen based on our initial literature search and availability of deep learning application papers enough to create a comprehensive review and curate insights within the water domain between 2018 to 2020. There were also other review articles partially covering the water domain and timeline (Shen, 2018).

After gathering the initial list totaling 1515 publications, each of them was briefly reviewed to determine whether they were in alignment with the scope of this study. All publications that are not research papers were excluded, namely vision papers, editorials and review papers. From the initial list of 1515 publications, 315 remained after this filtering step. These publications were filtered further to keep publications that met certain technical criteria. This step eliminated all publications that did not involve some form of deep artificial neural network in their pipeline of work.

After this step 129 publications remained and were included in our comprehensive review. The comprehensive review process consisted of manually reviewing the papers one by one to extract specific publication features, including: Architecture, Framework/Library/Programming Language, Dataset, Source Code Sharing, Reproducibility, Subfield: Deep Learning, Subfield: Environment, Summary. Each of the feature categories are described below:

- **Architecture** – The type of deep neural network architecture(s) employed in the study. This could be simply Artificial Neural Network (ANN) or more complex architectures like Generative Adversarial Networks (GAN) or Autoencoders (VAE).
- **Framework/Library** – This column serves as a survey within the field to understand the programming language and numeric computation library choices of researchers.
- **Dataset** – Whether the dataset(s) used in the study were collected specifically for the study, acquired from an authority resource, or previously existing standalone.
- **Source Code Sharing** – A boolean field indicating if the code of the study is open sourced and accessible by the public.
- **Reproducibility** – A boolean field indicating if the results of the study could be reproduced just by using the information provided in the manuscript.
- **Subfield: Deep Learning** – A classification of the machine learning task tackled in the paper. This field uses one of the following values; Regression, Classification, Sequence Prediction, Matrix Prediction, Unsupervised Learning and, Reinforcement Learning. Details of these are given in the next subsection where we describe deep learning practices.
- **Subfield: Environment** – A classification of the task carried out in an environmental field. This field uses one of the following values; Flood, Groundwater, Land Usage and Soil, Surface Water, Water Quality, Water Resources Management, Weather and, Others. Others include papers within the environmental field but do not exactly fit with other subfields we included here.
- **Summary** – The brief summary of the study.

These data for each publication reviewed herein are shared later in this section with figures and a table. Technical summaries of the papers reviewed are given in the Results section. Conclusions drawn from the acquired data are shared in the *Key Issues and Challenges* section.

**Deep Learning**

Deep Learning is a subfield of Machine Learning where a long-known algorithm, an artificial neural network (ANN) is used to map features into an output or a set of outputs. Formed by an input layer, intermediate hidden layers and an output layer, ANNs present an efficient way to learn linear and non-linear relationships between input and output pairs. Neural networks, when formed by many stacked layers, can represent complex features in later layers by using simpler representations formed by earlier layers in the network (Goodfellow et al., 2016). Each layer within an ANN comprises at least a neuron. ANN is a network of these neurons connected to each other with some weights and these neurons run specific functions, namely activation functions, mapping it's input to an output. Stacked on top of each other, the series of functions runs over the input of the network, translates the input to the output in the output layer. Typically, each neuron within a layer runs the same activation function and type of the layer is determined by this activation function. Network type is determined by the combination of layers used and how neurons are connected to each other within and between layers. The quintessential form of an ANN is the multilayer perceptron (MLP). An MLP contains at least a single hidden layer while each neuron within the network is connected to every neuron within the next layer. This architecture forms a fully connected neural network.

An activation function in a typical MLP multiplies the input by a weight and outputs it to the next neuron in line. In modern neural networks the common (Goodfellow et al., 2016) and recommended (Nair and Hinton, 2010) activation function is the Rectified Linear Unit (ReLU) function which introduces non-linearity to the network. A hidden layer that applies ReLU to the input could be referred as ReLU layer or an activation layer using ReLU. ANNs are generally trained using the backpropagation (BP) algorithm (Rummelhart et al., 1986) to fit the network to the set of input-output pairs using a loss function. A loss function or a cost function is used to determine how successful the mapping is done by a model. An easy and go-to loss function is mean squared error (MSE) which computes the difference between an output of the ANN and the ground truth for each input-output sample, squares the difference to avoid negative values and computes the mean error of all the samples. BP trains an ANN using the loss function by computing the gradient of the loss function with respect to the weights between neurons and updates the weights to better fit the data samples in order to minimize the loss function.

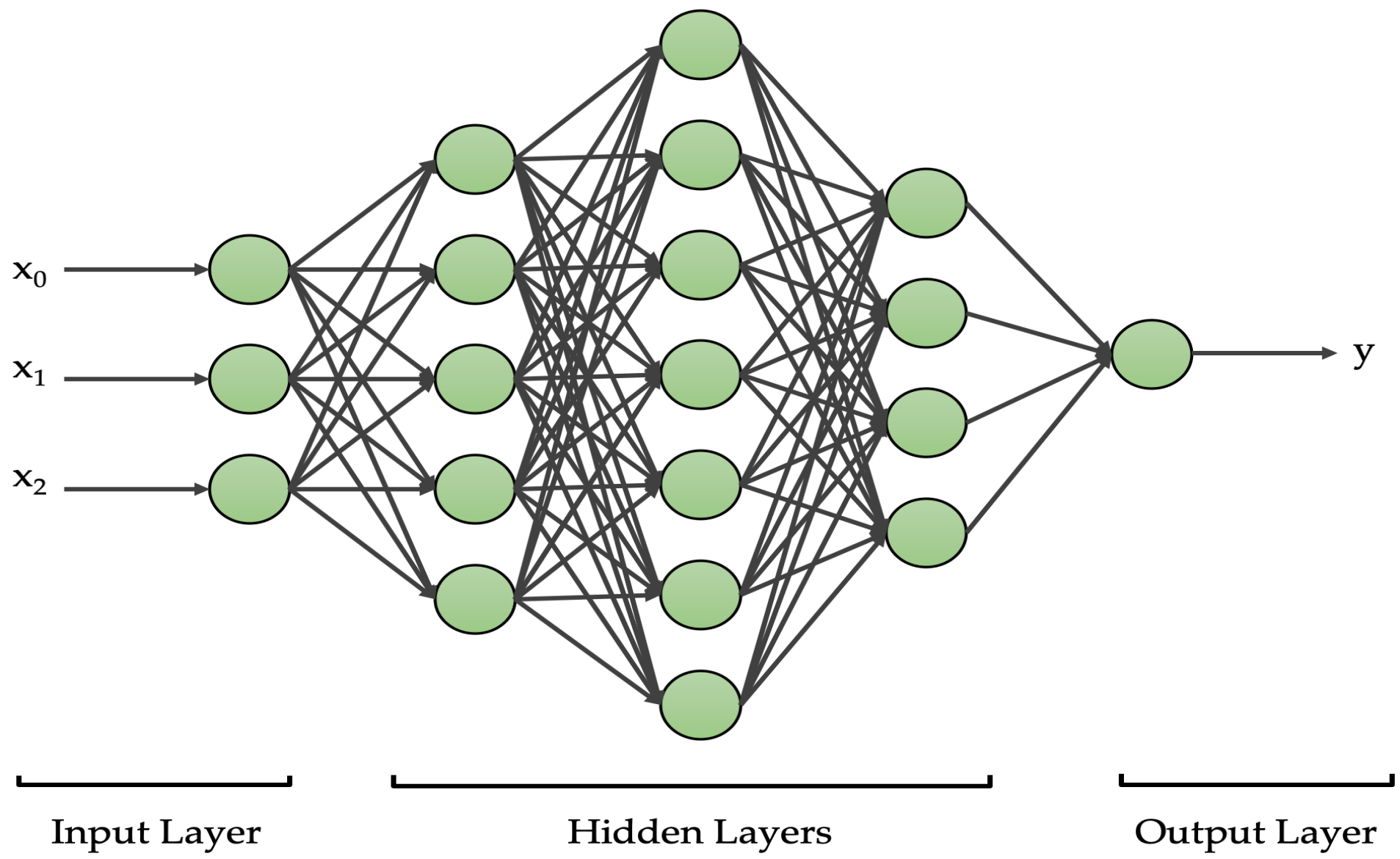

Figure 1. A densely connected artificial neural network architecture

ANNs are not a new concept, considering primitive versions were around in the 1940s (Goodfellow et al., 2016). Instead, ANNs attracted attention of researchers from various scientific disciplines when it became clear that they are extremely powerful in capturing representations in the data and with advances in graphics processing units (GPUs) which enable extremely fast matrix operations (Goodfellow et al., 2016). In this way, a neural network architecture that previously was infeasible to utilize due to time complexity as the number of hidden layers increased, could be used in training on complex datasets learning representations.

For each of the reviewed papers we identify the machine learning subfield, which correspond to task types. Task types are defined by the output form of a network. We consider the following as distinct task types: Regression, Classification, Sequence Prediction, Matrix Prediction, Unsupervised Learning, and Reinforcement Learning. These task types are briefly discussed below:

- **Regression** – A machine learning task that predicts a continuous value. Forecasting the next measurement from a series of measurements would be an example of a regression task.
- **Classification** – When the aim of a model is to predict whether the set of inputs can be categorized into some classes, the task is a classification task. The number of classes are not limited. Predicting whether the next day will be a dry or wet one is a classification task.
- **Sequence Prediction** – Regression of a sequence of numeric values or a vector. Forecasting the next 24 hours of measurements for a stream sensor is a sequence prediction task.
- **Matrix Prediction** – Regression of a matrix of numeric values. Forecasting the next precipitation measurements for a rectangular region would be a matrix prediction task. Each of the values in the predicted matrix would be the precipitation value of a subregion within the actual region.
- **Unsupervised Learning** – A learning task applying competitive learning instead of error correction learning like previous task types. Decreasing dimension of a high-dimensional hydrological input data to lower-dimension visualizable data would be an example.
- **Reinforcement Learning** – A learning task where the output is unspecified. In a reinforcement learning task, the algorithm tries to find the optimal solution for any given input using a reward/penalty policy and a try-error mechanism. An example of a reinforcement learning task is an AI model that learns when to release water from a dam.

**Architectures**

This subsection summarizes cornerstone neural network architectures used by papers reviewed in this study. Also, some ANN concepts are briefly discussed. For further understanding of these architectures, we refer readers to the cited works.

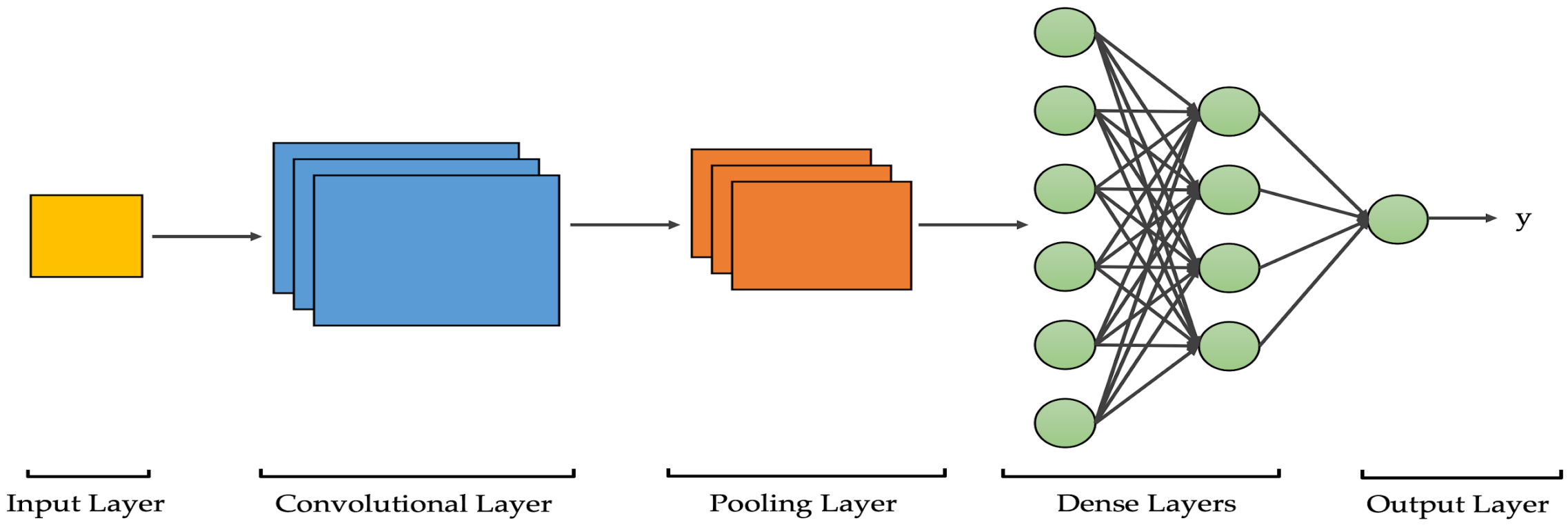

Figure 2. A basic convolutional neural network structure for image classification

## Convolutional Neural Networks (CNNs)

A CNN (LeCun, 1989) or a ConvNet comprises at least one convolutional layer, which methodologically expects a 3D tensor as the input. A convolutional layer applies multiple cascaded convolution kernels to extract intricate knowledge from the input. For example, a CNN for an RGB image tensor with a shape of *image width* x *image height* x 3 would have a convolutional layer which applies 3 different convolution operations with 3 separate kernels to each of the color channel matrices. Using a convolution kernel matrix, a convolutional layer that processes an image as such can extract 2D positional information from images, such as understanding objects that are close to each other.

To make a neural network with a convolutional layer cognize non-linear correlations between the input and output along with linear correlations, one needs to introduce non-linearity to the network via an activation layer. Typically, that is done by using the Rectified Linear Unit (ReLU) as the activation function following the convolution layer. Another common layer used within a CNN is a pooling layer. A pooling layer is used to reduce the size of the input while keeping the positional knowledge intact. A frequently used pooling method within CNN literature is Max Pooling (Zhou and Chellappa, 1988). This sample-based discretization moves the most important learnt features to subsequent layers while reducing the size. Consequently, CNNs make good architectures for deep learning tasks with images or image-like objects as inputs. This ability of CNNs makes way for various breakthroughs in the fields of object detection, super-resolution, image classification, and computer vision.

## Generative Adversarial Networks (GANs)

GANs (Goodfellow et al., 2014) consist of two seemingly separate CNNs working in unison and competing in a min-max game. One of these CNNs, the generator, aims to generate fake examples out of a dataset while the other, discriminator, aims to reveal whether its input is fake or not. Since they try to beat each other, it causes them to get better over time in both generating fake outputs and discriminating fake from real.

GANs are initially used as generative models, as in randomly generating new samples from a dataset to appear as if they are from the originating dataset when visualized (Gautam et al., 2020). They achieve this goal by mapping random noise to real samples from the given dataset, and then they generate new instances from new random noise tensors. Despite their success at generation, GANs are also capable of learning translation tasks such as super-resolution (Demiray et al., 2020) or image to image translation (Isola et al., 2017).

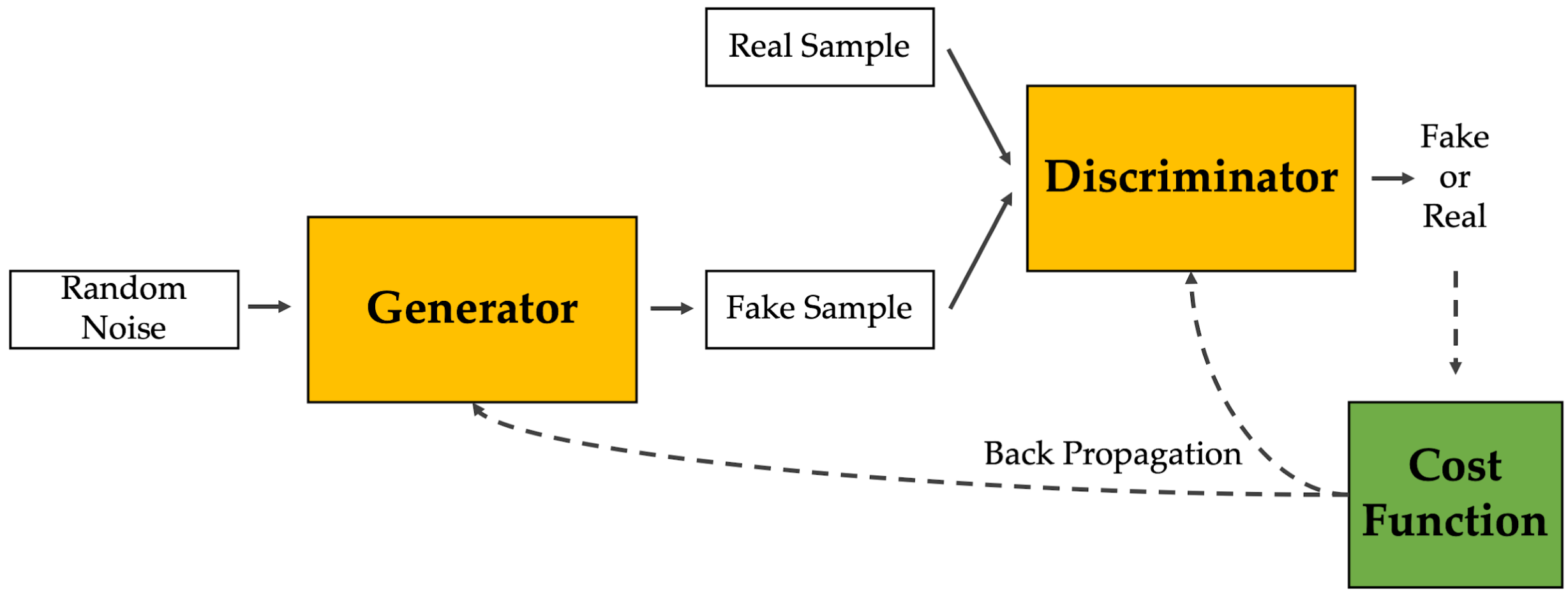

Figure 3. Overview of generative adversarial networks

**Recurrent Neural Networks (RNNs)**

RNNs (Pollack, 1990) are a type of artificial neural network that includes a recurrent layer. The difference of a recurrent layer from a regular fully-connected hidden layer is that neurons within a recurrent layer could be connected to each other as well. In other words, the output of a neuron is conveyed both to the neuron(s) within the next layer and to the next neuron within the same layer. Using this mechanism, RNNs can carry information learned within a neuron to the next neuron in the same layer.

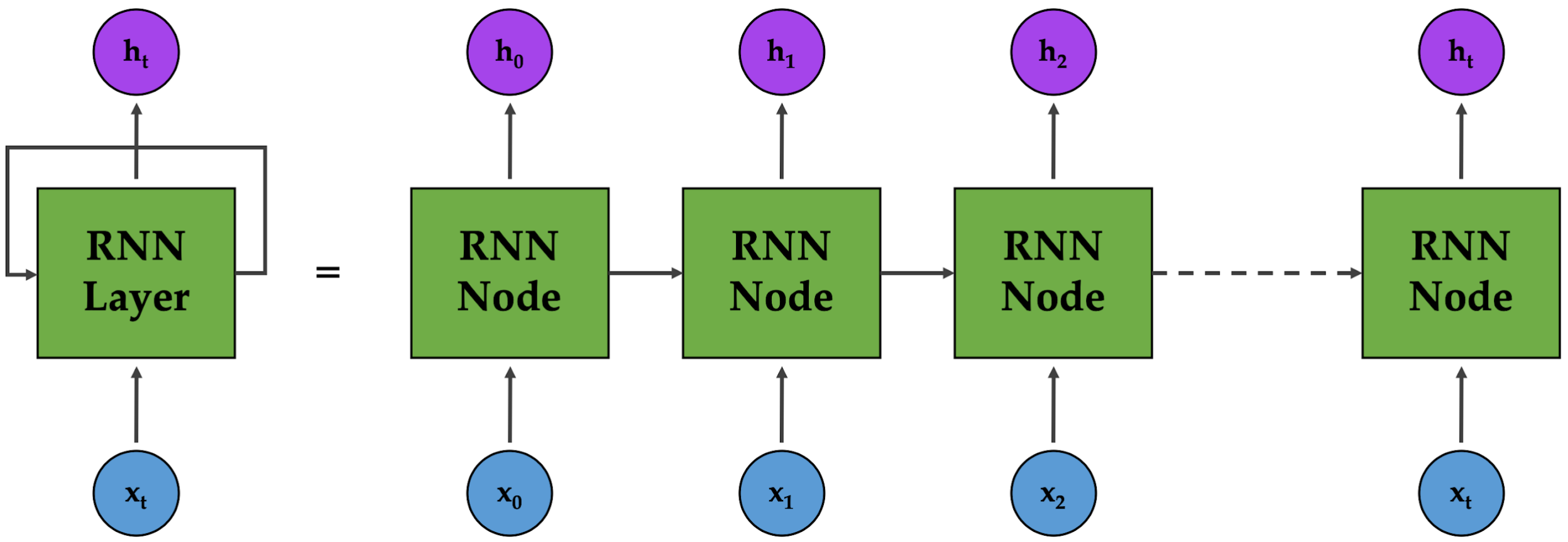

Figure 4. Connections of a Recurrent layer

This procedure becomes convenient when the data to be fed to the network is in sequential nature such as a time-series data or a text. When training a model over a data sample like a text to extract the meaning, most of the time, the beginning of the text could change the meaning that is to be extracted from the end of the text. RNNs aim to keep the information gained from earlier parts of a data sample in the memory and move it to the later parts of the same data sample to ensure better knowledge discovery (Goodfellow et al., 2016). A simple RNN implementation lacks the practicality in long sequences, such as long paragraphs, as it is common to encounter the *vanishing gradient problem* while training (Bengio et al., 1993). With the vanishing gradient problem, the gradients of the loss function get extremely high in some cases during training and consequently make the training process and the trained network deficient (Goodfellow et al., 2016). More complex RNN implementations like Long Short-Term Memory (LSTM) (Hochreiter and Schmidhuber, 1997) Networks or Gated Recurrent Unit (GRU) Networks (Cho et al., 2014) solve this problem but have greater computational complexity. Various RNN structures can be used for tasks that somewhat rely on sequential understanding of datasets such as language modeling (Sundermeyer et al., 2012), text classification (Yin et al., 2017; Sit et al., 2019) and time-series forecasting (Xiang and Demir, 2020; Sit and Demir 2019).

**Long Short-Term Memory (LSTM) Networks**

LSTM (Hochreiter and Schmidhuber, 1997) Networks are developed for longer short term memory life over the input, paving the way for more efficient but more resource intensive training over datasets consisting of sequential samples. Instead of an activation function producing one output and carrying the output to immediate neurons both in the next layer and the same layer, LSTM neurons produce two different values yielded by a series of activations and operations. While both outputs are kept within the LSTM layer to keep track of things learnt over the past part of the sequence, one of the outputs is transferred to the next layer (Figure 5).

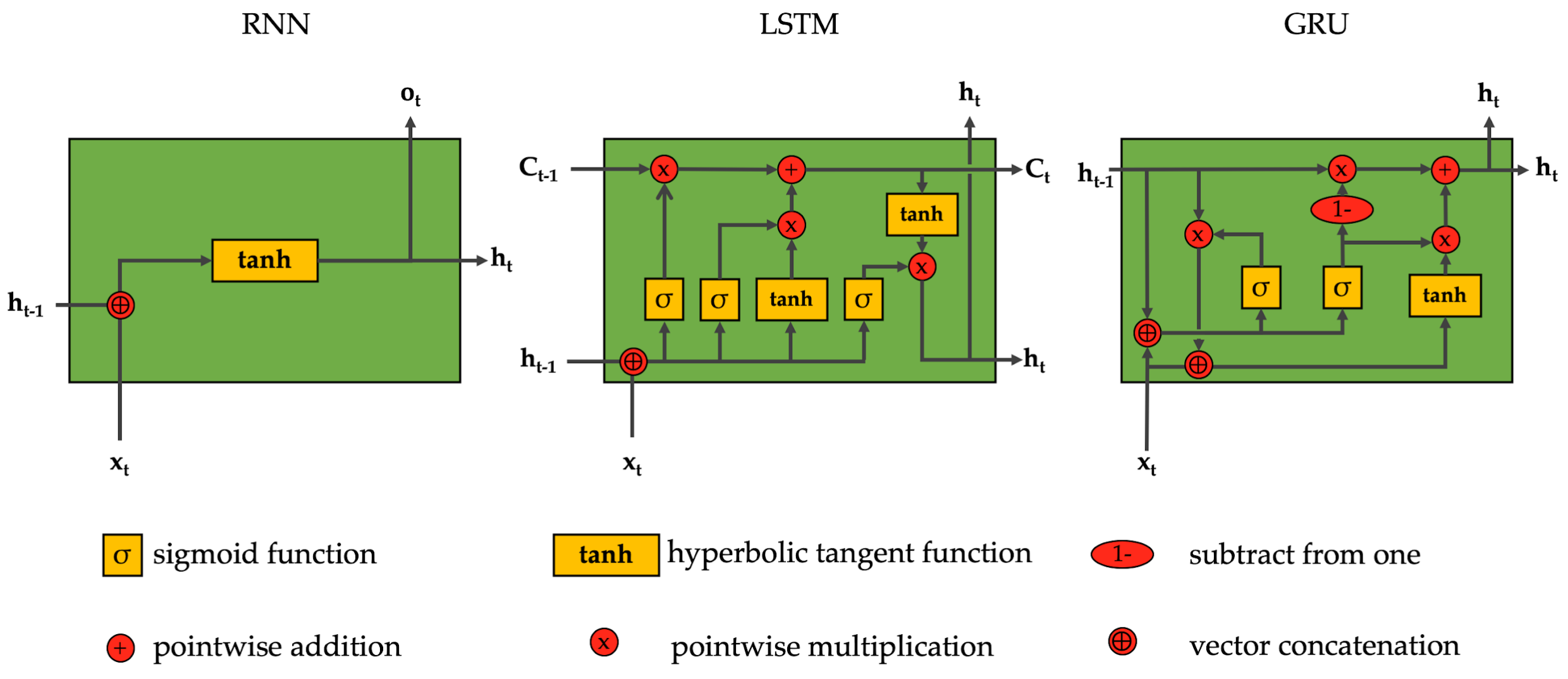

Figure 5. Computation-wise comparison of RNN, LSTM and GRU nodes

**Gated Recurrent Unit (GRU) Networks**

Although LSTM Networks, most of the time, solve the vanishing gradients problem and helped many breakthroughs within the fields of Natural Language Processing and time-series prediction, their time complexity emerges as a downside. GRU (Cho et al., 2014) networks reduce the complexity while keeping the efficacy intact. Similar to a simple RNN neuron, a GRU neuron produces only one output after a series of computations and uses the same output to convey important features learnt to both the next layer and the next neuron within the same layer.

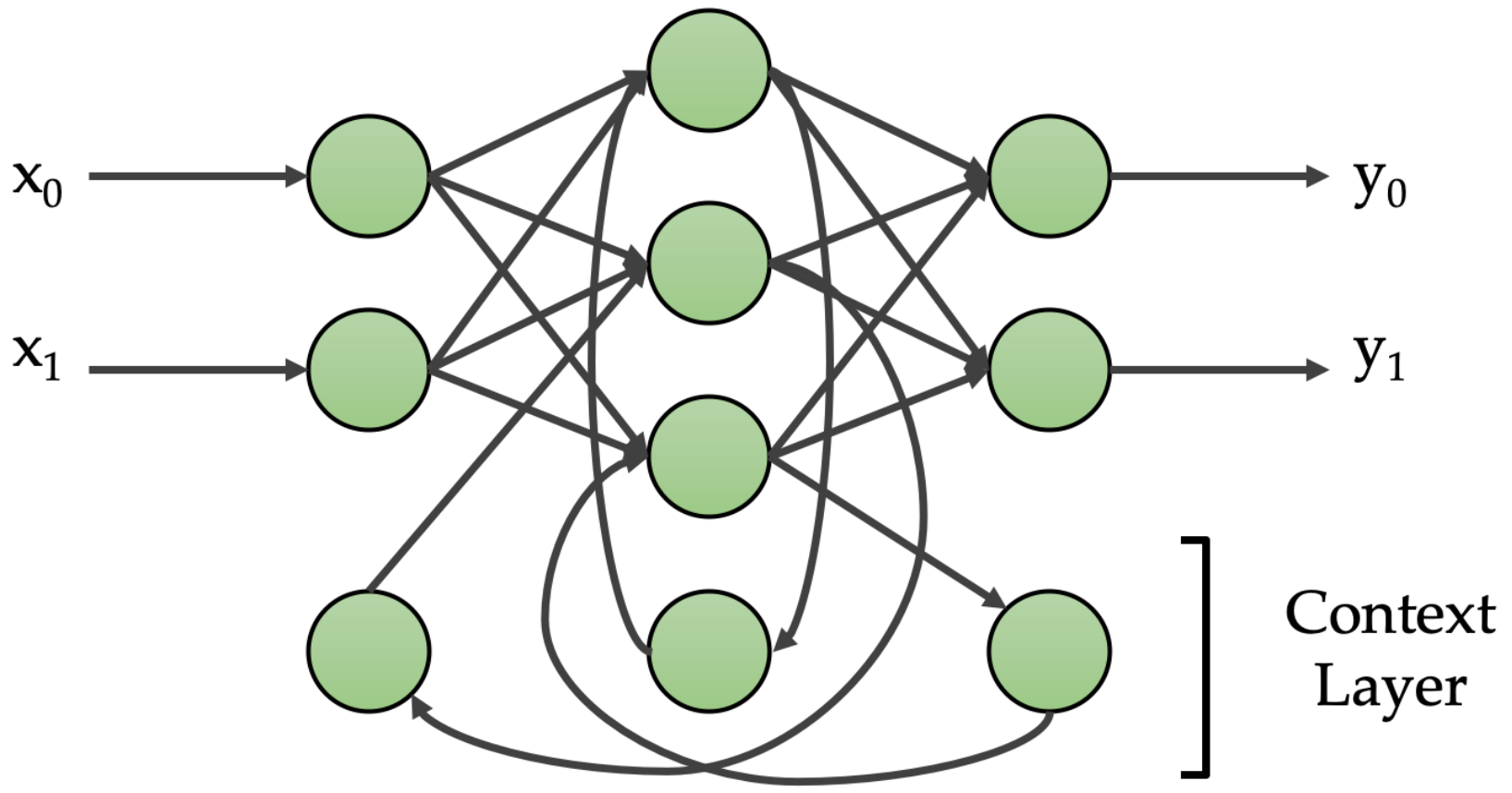

Figure 6. An Elman Network architecture with two input and two output neurons

**Nonlinear Autoregressive (NAR) Models**
A NAR model is not necessarily a neural network model but a model that is used for time-series prediction, taking into account both current and previous samples from a time-series to map input sequence to outputs. A NAR model needs a nonlinear function such as a polynomial function or a neural network to perform training. If a neural network is used, a NAR network would classify as an RNN based on the fact that it utilizes sequential complexion of the given input. Papers reviewed within this study that employ NAR, thus, implement a neural network as the function in their proposed models. There are many NAR variations and one that deserves mention, due to cardinality of papers reviewed in this study which employ it, is the Nonlinear Autoregressive Exogenous Model (NARX) (Lin et al., 1996). NARX is an RNN implementation that takes advantage of exogenous inputs, hence the name.

**Elman Network (ENN)**
An Elman Network (Elman, 1990) is yet another RNN implementation that has three layers, only one being a hidden layer. The hidden layer of the Elman Network is connected to a set of neurons called context units. In each iteration after the first one over the network, the state of the hidden layer is copied into the context units. Thus, the state of the network for the previous sample in the data stream is kept in the network each time to be used in next iterations. An Elman Network can train over sequential datasets better than a regular ANN due to this mechanism acting like a memory.

**Autoencoders (AE)**
Autoencoders (Rummelhart et al., 1985) are neural networks that are used to reduce the dimensionality of datasets. They are implemented in an unsupervised fashion to generate only a representation of the dataset within their hidden layer neurons, also called the latent vector. Taking the same set of values for both input and output of the network, an AE learns to reduce a dataset into a representation state and additionally learns how to reconstruct the data sample to its original form from the learned representation.

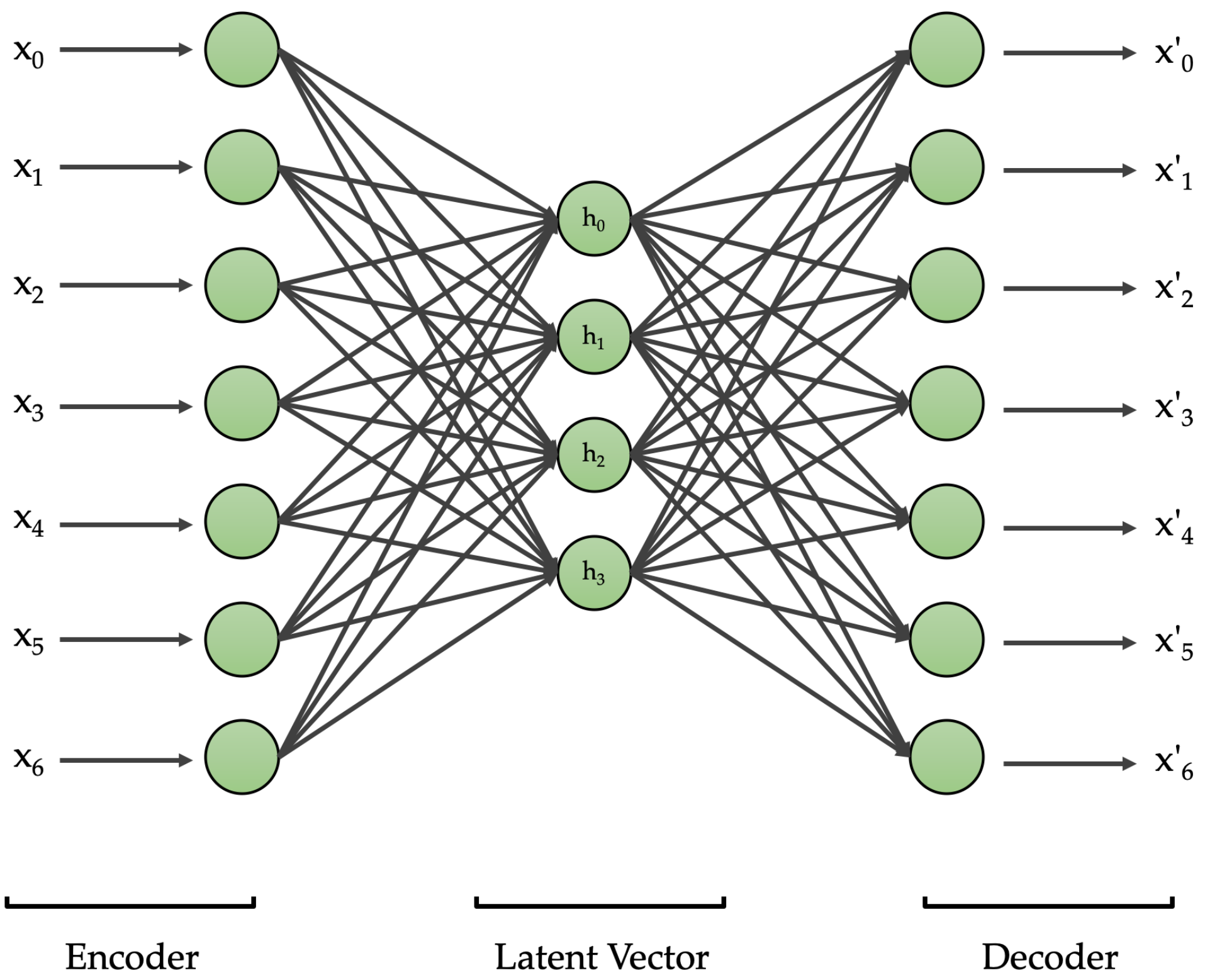

Figure 7. Visualization of an Autoencoder network

**Deep Q Networks (DQN)**
DQN (Mnih et al., 2013) is a reinforcement learning algorithm in which there is a predetermined rewarding policy. The DQN works similarly to the Q-Learning algorithm which works by providing a medium for the agent and the environment and tries to maximize the rewards the agent gets through its actions in the environment. A DQN differentiates from the conventional Q-Learning algorithm with how it generalizes. In other words, the agent in the Q-Learning algorithm cannot estimate the outcome from action-environment pairs it didn't see before, while in DQN the agent is able to produce a reinforced action.

**Restricted Boltzmann Machines (RBMs) and Deep Belief Networks (DBNs)**
RBMs (Hinton, 2002) present two-layer stochastic generative models that are in the shape of bipartite graphs. RBMs form the building block for deep belief networks (DBN) (Hinton, 2009), but they can also be used as standalone models. RBMs were initially used in unsupervised tasks, they also enable the user to tackle classification and regression tasks by implementing them within other networks. The paper reviewed in this study that employs RBMs uses it in a setting where RBMs are followed by a set of fully-connected layers in order to perform a classification task. Formed by stacked RBMs, DBNs, similarly to RBMs, could also be used to tackle many types of tasks by training unsupervised beforehand. The bipartite connections of RBMs are altered when they are stacked onto each other to form a DBN (Figure 8). Only the top two layers of a DBN have bipartite connections while the rest of the layers have one-way connections with each other. Though RBMs are useful for many tasks, their use in research is decreasing in the machine learning field as researchers adopt newer architectures that could be utilized in the same fashion such as AE, DBNs and GANs.

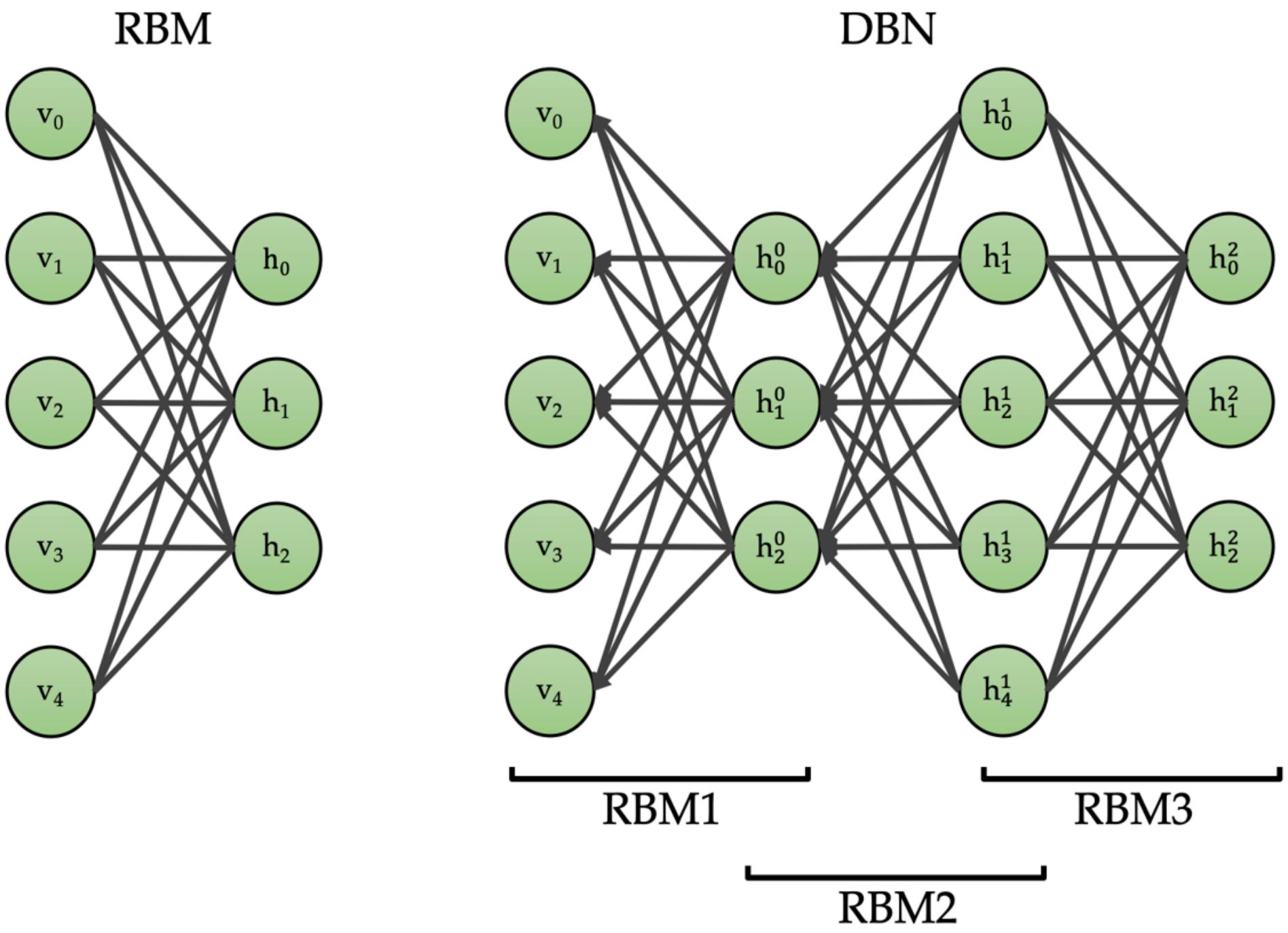

Figure 8. RBM and DBN examples

**Extreme Learning Machines (ELM)**
Extreme Learning Machines (Huang et al., 2006) are three-layer neural networks in which the weights connecting to the second layer from the input layer are randomized, and the weights connecting to the third are trained. ELM networks have been criticized for being unoriginal (Wang and Wan, 2008).

**Summary of Articles**
This subsection presents visual summaries of the reviewed papers as well as a summary table extracted information from each paper that received comprehensive review. As previously stated, information from a total of 129 papers are given, however data points in shared figures do not always equal to this value depending on the data column. This is due to some data columns storing multiple values at the same time, and in some cases, papers not including the relevant data.

Among the reviewed papers, as in the deep learning literature, most used architectures are CNNs and LSTMs (Figure 9). We explain this aspect with their respective success in matrix prediction and sequence prediction, tasks that have high importance in hydrologic modelling. One confounding thing is that even though LSTM networks were vastly employed, one architecture that yields similar performance, GRU networks did not find significant usage in the field. Additionally, we found it surprising that even though most of the studies reviewed here tackles tasks involving sequential data, Transformers were not employed by any of the studies we reviewed. It should be noted that a Transformer is a neural network architecture that is vastly used in the field of natural language processing, which is another field that focuses on sequential data.

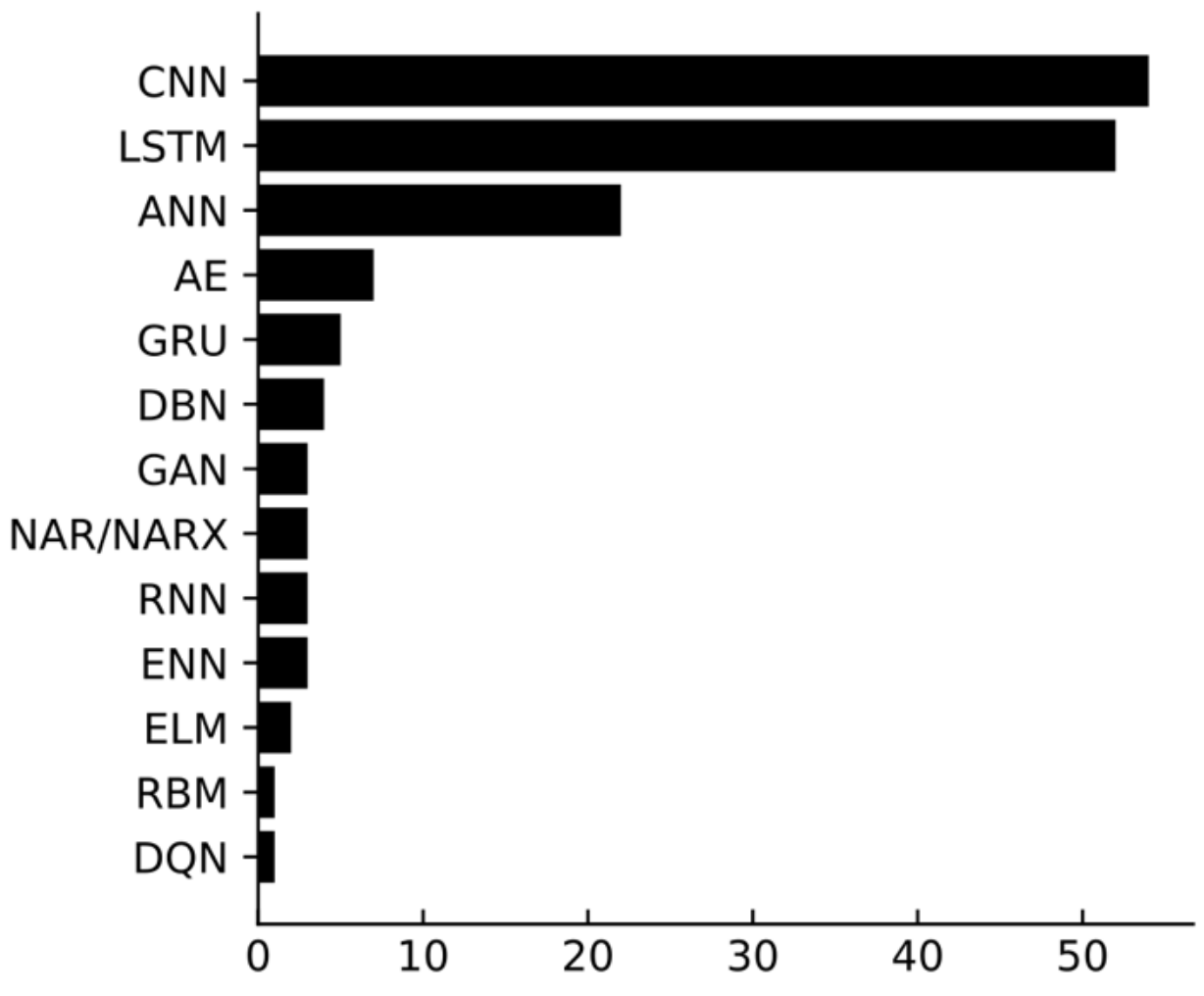

Figure 9. Architectures described in the study and their usage in reviewed papers

As the utilization of deep learning in the water field in a broad sense increases over time (Figure 10), annual usage of the deep neural network architectures increases. Figure 11 shows the change in usage through March 2020 and also presents projections for the rest of the year built on top of the number of publications through March 2020. We expect to see growth in the usage of neural network architectures that have been widely used in other disciplines but not in the water field like DQNs and GANs.

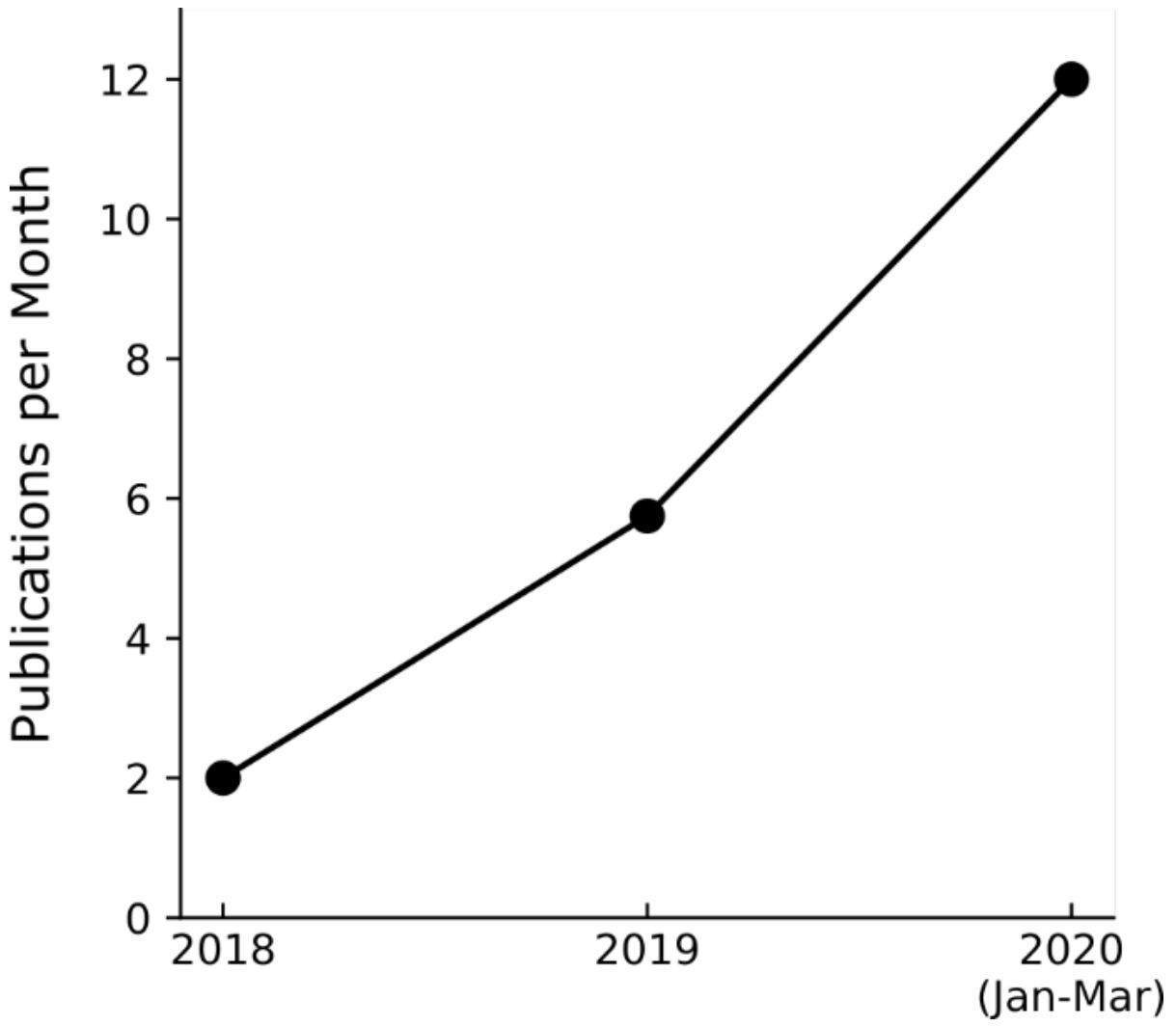

Figure 10. Average number of papers published each month during 2018, 2019 and 2020 January to March

The datasets used in publications reviewed typically are datasets acquired from authorities or governmental agencies (Figure 12). Even though in deep learning literature, the dataset acquisition primarily done by using previously existing datasets, in the water field, on the contrary, this does not seem to be the case. Code accessibility of the papers is another aspect of the studies published in the water field that differentiates it from the deep learning field in general. Although open-sourced models are widely expected from deep learning researchers, open-sourcing the software built for a study is unusual for a publication in the water field if not rare, as it can be seen in Table 1. Cumulating from both the data acquisition type and code accessibility, reproducing the outcomes of a papers does not seem to be an easy task for the authorities and other researchers in the field.

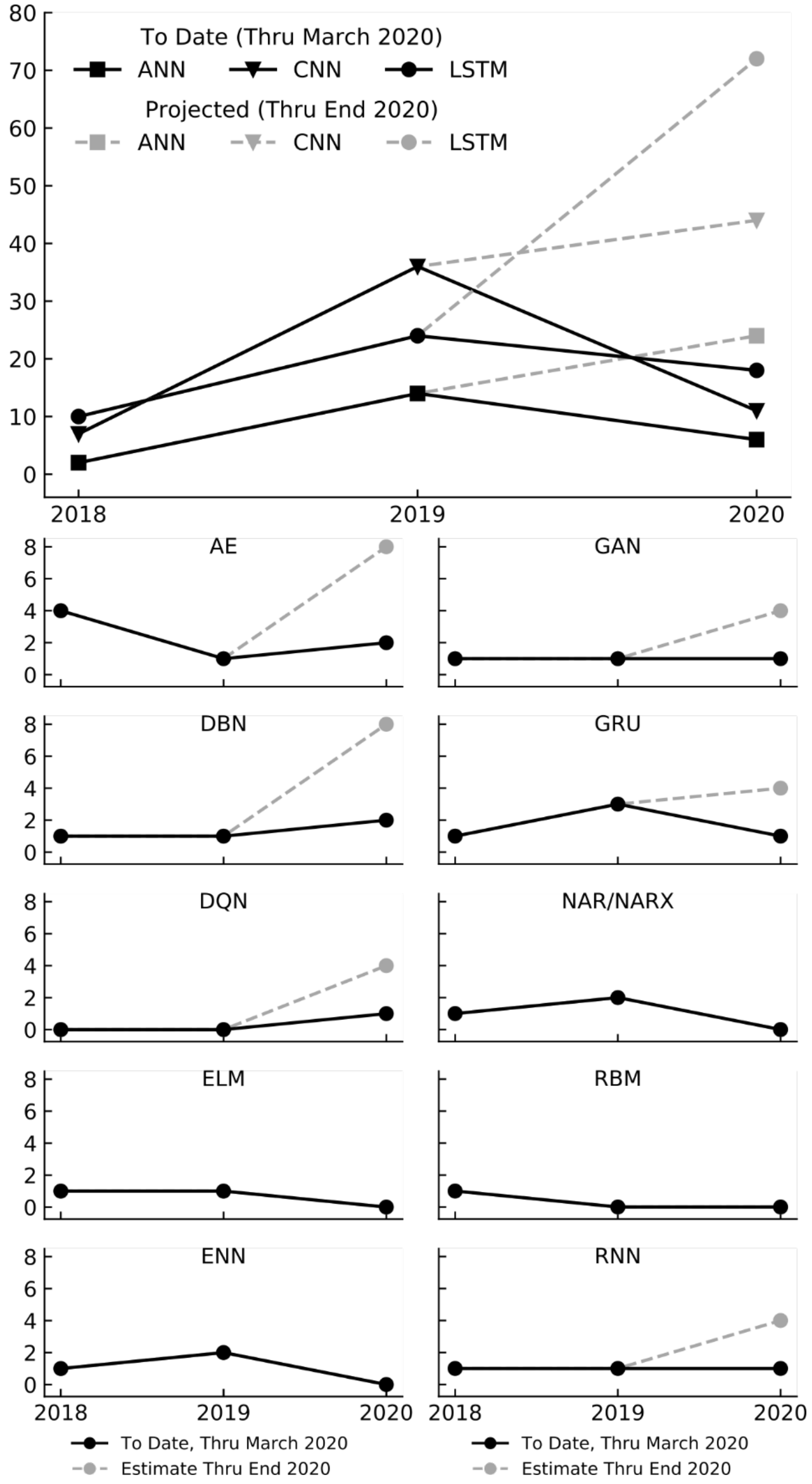

Figure 11. Architectures described and their annual usage in reviewed papers

Owing to the fact that the field of hydrology vastly relies on sequential data, most studies seem to work on sequence prediction and regression tasks (Figure 13). This phenomenon also might be linked with the fact that most of the studies reviewed were classified in Flood subdomain (Figure 14). Figure 15 summarizes the usage of numerical computing frameworks. Even though TensorFlow seems to be the first choice among water domain researchers, it should be noted that most of the usage comes from Keras, the second most used framework, which typically works on top of TensorFlow by providing a higher-level interface. Thus, our inference is that Keras is the most used deep learning framework within the water field. In contrast, libraries like PyTorch that is highly endorsed in deep learning literature find a smaller place to themselves.

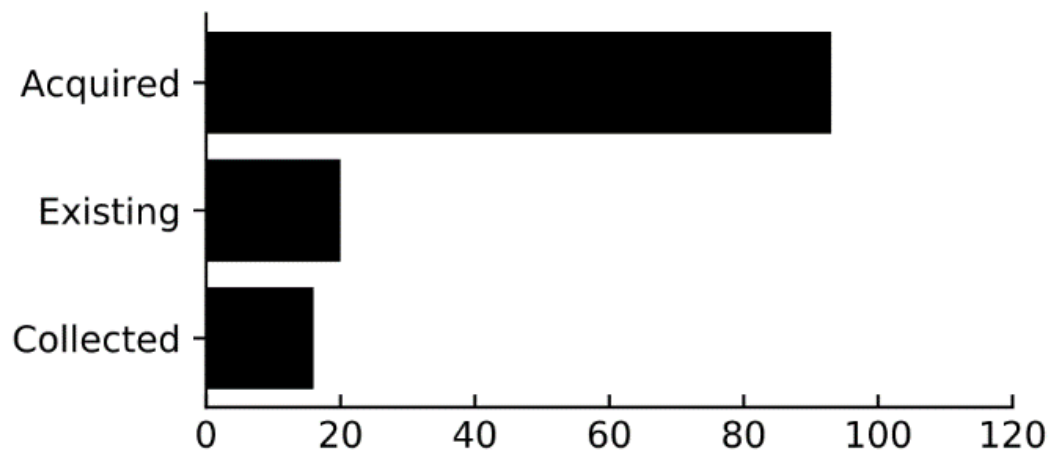

Figure 12. Distribution of dataset acquisition types in reviewed papers

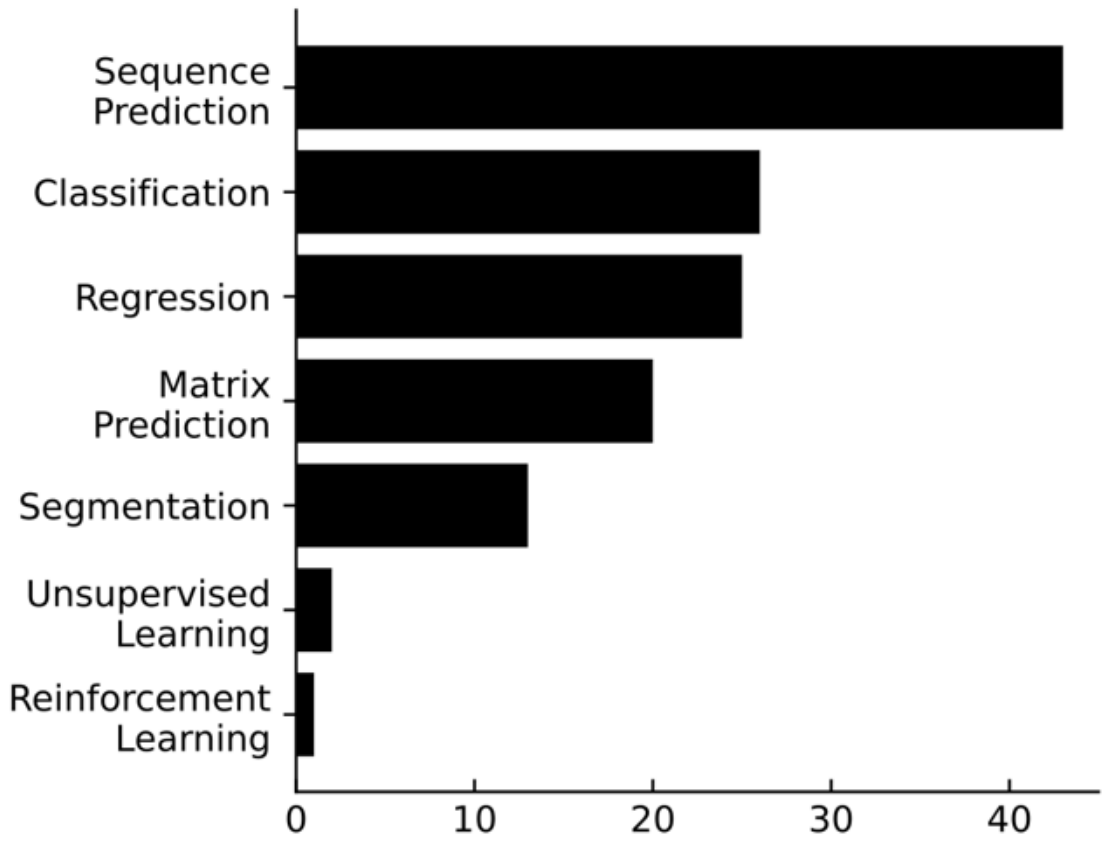

Figure 13. Histogram of machine learning task types studied in reviewed papers

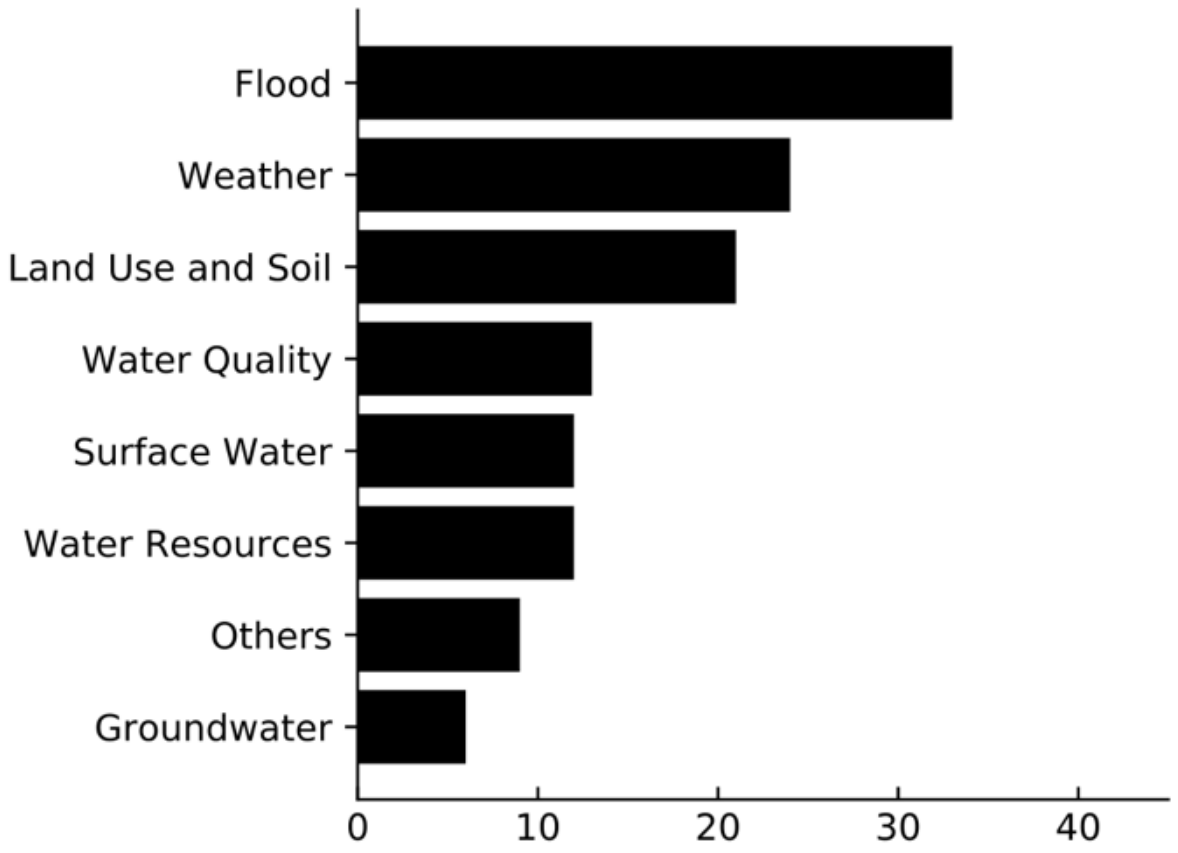

Figure 14. Histogram of Water domain subfields studied in reviewed papers

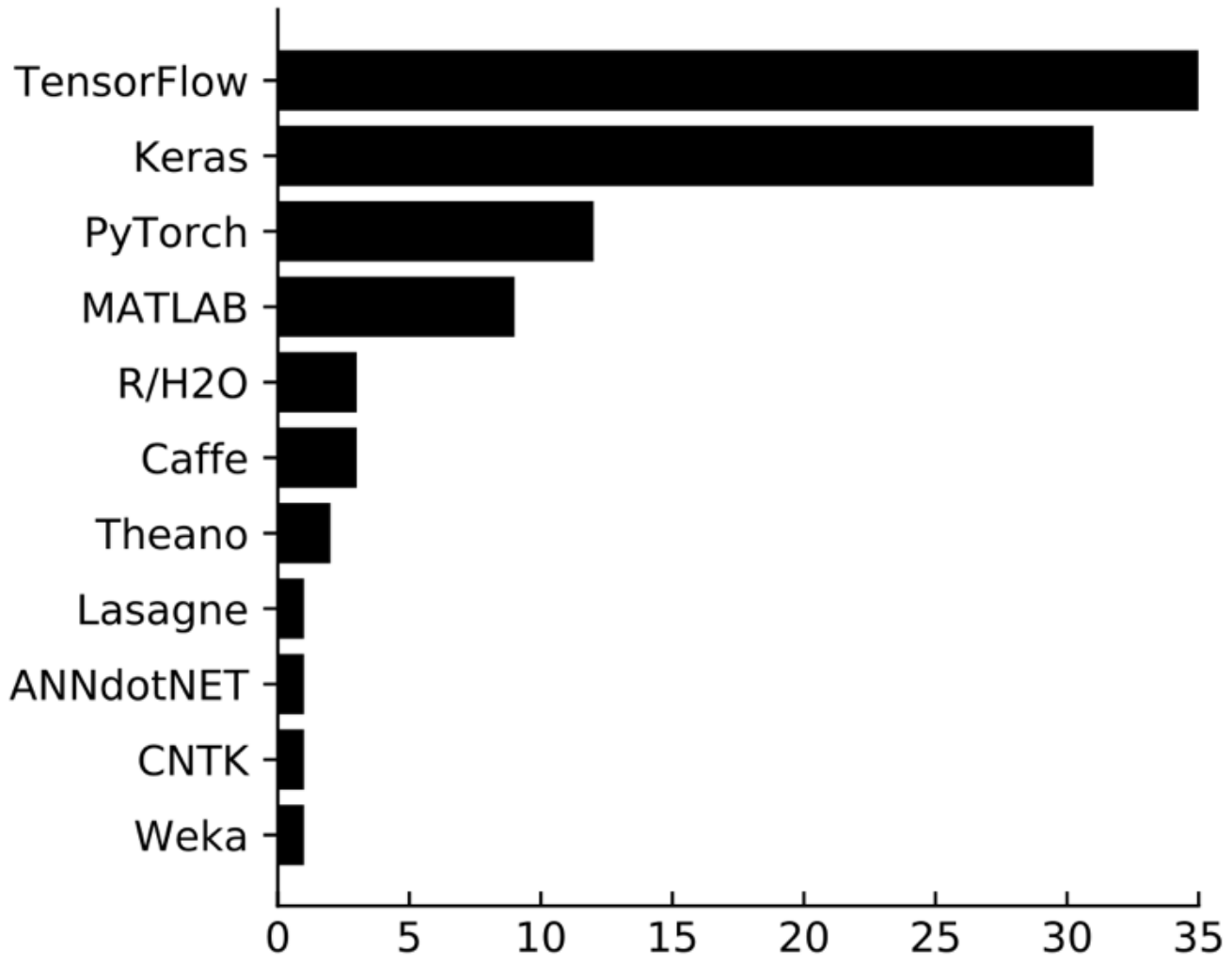

Figure 15. ML frameworks described in the study and their usage in reviewed papers

Table 1. Reviewed papers with curated data points

| Paper | Network Type | Framework | Dataset | Open Source | Reproducible | DL Task | Water Field |
|---|---|---|---|---|---|---|---|
| Yuan et al., 2018 | LSTM | MATLAB | Acquired | No | No | Sequence Prediction | Flood |
| Di Zhang et al., 2018 | LSTM | - | Acquired | No | No | Sequence Prediction | Flood |
| Kratzert et al., 2018 | LSTM | Keras, TensorFlow | Existing | No | Yes | Sequence Prediction | Flood |
| X. He et al., 2019 | ANN, DBN | - | Acquired | No | Yes | Sequence Prediction | Flood |
| Hu et al., 2019 | LSTM | - | Acquired | No | No | Regression | Flood |
| J.H. Wang et al., 2019 | CNN | - | Acquired | No | No | Sequence Prediction | Flood |
| S. Yang et al., 2019 | NAR, LSTM | Keras, TensorFlow | Acquired | No | No | Regression | Flood |
| Sankaranarayanan et al., 2019 | ANN | Keras | Acquired | No | No | Classification | Flood |
| Ni et al., 2019 | LSTM, CNN | - | Acquired | No | No | Sequence Prediction | Flood, Weather |
| Bai et al., 2019 | LSTM, AE | - | Acquired | No | No | Sequence Prediction | Flood |
| T. Yang et al., 2019 | LSTM | - | Acquired | No | No | Regression | Flood |
| Bhola et al., 2019 | CNN | - | Collected | No | No | Segmentation | Flood |
| Damavandi et al., 2019 | LSTM | Keras, TensorFlow | Acquired | No | Yes | Sequence Prediction | Flood |
| Moy de Vitry et al., 2019 | CNN | TensorFlow | Collected | Yes | Yes | Matrix Prediction | Flood |
| Worland et al., 2019 | ANN | Keras, TensorFlow | Acquired | No | No | Regression | Flood |
| Kratzert et al., 2019a | LSTM | PyTorch | Acquired | Yes | Yes | Sequence Prediction | Flood |
| Kumar et al., 2019 | RNN, LSTM | Keras | Acquired | No | Yes | Sequence Prediction | Flood |
| Wan et al., 2019 | ENN | - | Acquired | No | No | Sequence Prediction | Flood |
| Qin et al., 2019 | LSTM | TensorFlow | Acquired | No | No | Sequence Prediction | Flood |
| Kratzert et al., 2019b | LSTM | PyTorch | Existing | Yes | Yes | Regression | Flood |
| D.T. Bui et al., 2020 | ANN | MATLAB | Acquired | No | No | Classification | Flood |
| Nguyen and Bae, 2020 | LSTM | TensorFlow | Acquired | No | No | Sequence Prediction | Flood |
| Q.T. Bui et al., 2020 | ANN | - | Acquired | No | No | Classification | Flood |
| Kabir et al., 2020 | DBN | TensorFlow | Acquired | No | No | Sequence Prediction | Flood |
| Kao et al., 2020 | LSTM | Keras | Acquired | No | No | Sequence Prediction | Flood |
| Xiang et al., 2020 | LSTM | Keras, TensorFlow | Acquired | No | No | Sequence Prediction | Flood |
| Zuo et al., 2020 | LSTM | Matlab, TensorFlow | Acquired | Yes | Yes | Sequence Prediction | Flood |
| Ren et al., 2020 | ANN, LSTM, GRU | - | Acquired | No | No | Sequence Prediction | Flood |
| Shuang Zhu et al., 2020 | LSTM | Keras | Acquired | No | Yes | Sequence Prediction | Flood |
| Y. Wang et al., 2020 | CNN | Keras | Acquired | No | No | Classification | Flood |
| Laloy et al., 2018 | GAN | Lasagne, Theano | Existing | No | No | Matrix Prediction | Groundwater |
| N. Wang et al., 2020 | ANN | - | Collected | Yes | No | Matrix Prediction | Groundwater |
| Santos et al., 2020 | CNN | Keras | Existing | Yes | Yes | Matrix Prediction | Groundwater |
| Mo et al., 2019a | CNN | PyTorch | Collected | Yes | Yes | Matrix Prediction | Groundwater |
| Mo et al., 2019b | CNN | PyTorch | Collected | Yes | Yes | Matrix Prediction | Groundwater |
| A.Y. Sun et al., 2019 | CNN | Keras, TensorFlow | Acquired | No | No | Sequence Prediction | Groundwater |
| Jeong and Park, 2019 | NARX, LSTM, GRU | TensorFlow | Acquired | No | No | Sequence Prediction | Groundwater |
| Zhou et al., 2020 | CNN | PyTorch | Acquired | No | No | Regression | Groundwater |
| Jeong et al., 2020 | LSTM | TensorFlow | Acquired | No | No | Sequence Prediction | Groundwater |
| C. Zhang et al., 2018 | CNN | Keras, TensorFlow | Collected | No | No | Segmentation | Land Use and Soil |
| X. Zhang et al., 2018 | LSTM | TensorFlow | Acquired | No | No | Regression | Land Use and Soil |
| Cao et al., 2018 | CNN, ELM | - | Existing | No | No | Segmentation | Land Use and Soil |
| Zeng et al., 2018 | ANN | - | Acquired | No | No | Regression | Land Use and Soil |
| Reddy et al., 2018 | LSTM | - | Acquired | No | Yes | Sequence Prediction | Land Use and Soil |
| Fu et al., 2018 | CNN | - | Acquired | No | No | Classification | Land Use and Soil |
| Jiang, 2018 | AE | - | Acquired | No | No | Classification | Land Use and Soil |
| Shen et al., 2019 | ANN | R/H2O | Acquired | No | No | Regression | Land Use and Soil |
| Jin et al., 2019 | CNN | TensorFlow | Acquired | No | No | Segmentation | Land Use and Soil |
| Persello et al., 2019 | CNN | - | Acquired | No | No | Segmentation | Land Use and Soil |
| Kroupi et al., 2019 | CNN | TensorFlow | Existing | No | No | Classification | Land Use and Soil |
| Z. Sun et al., 2019 | CNN | PyTorch | Acquired | No | No | Classification | Land Use and Soil |
| Meng et al., 2019 | CNN | - | Acquired | No | No | Classification | Land Use and Soil |
| Kopp et al., 2019 | CNN | - | Collected | No | No | Segmentation | Land Use and |
| | | | | | | | Soil |
| Jiang et al., 2019 | CNN | - | Acquired | No | Yes | Matrix Prediction | Land Use and Soil |
| Bhosle and Musande, 2019 | CNN | TensorFlow | Existing | No | No | Classification | Land Use and Soil |
| C. Zhang et al., 2019 | ANN, CNN | - | Existing | No | No | Matrix Prediction | Land Use and Soil |
| J. Wang et al., 2020 | DBN | - | Acquired | No | No | Matrix Prediction, Regression | Land Use and Soil |
| Nam and Wang, 2020 | AE | R/H2O | Acquired | No | No | Classification | Land Use and Soil |
| O'Neil et al., 2020 | CNN | TensorFlow | Acquired | Yes | Yes | Classification | Land Use and Soil |
| Yang et al., 2020 | CNN | Keras | Acquired | No | Yes | Segmentation | Land Use and Soil |
| Abdi et al., 2018 | CNN | - | Existing | No | Yes | Classification | Others |
| Li et al., 2018 | CNN, AE | PyTorch | Acquired | No | Yes | Segmentation | Others |
| Rohmat et al., 2019 | ANN | MATLAB | Acquired | No | No | Regression | Others |
| Amirkolaee and Arefi, 2019 | CNN | MATLAB | Existing | No | Yes | Matrix Prediction | Others |
| C. Wang et al., 2019 | CNN | Keras | Existing | No | No | Classification | Others |
| Kylili et al., 2019 | CNN | - | Collected | No | No | Classification | Others |
| Kang, 2019 | CNN | Keras, TensorFlow | Acquired | No | No | Classification | Others |
| Haklidir and Haklidir, 2019 | ANN | - | Acquired | No | Yes | Regression | Others |
| Kim et al., 2020 | GAN, AE | - | Acquired | No | No | Unsupervised Learning | Others |
| J. Zhang et al., 2018 | LSTM | Theano | Acquired | No | No | Regression | Surface Water |
| Yongqi Liu et al., 2019 | ANN | Keras, TensorFlow | Acquired | No | Yes | Sequence Prediction | Surface Water |
| C. Xiao et al., 2019 | CNN, LSTM | Keras, TensorFlow | Acquired | No | No | Matrix Prediction | Surface Water |
| Read et al., 2019 | LSTM | - | Acquired | No | No | Regression | Surface Water |
| Yansong Liu et al., 2019 | CNN | Caffe | Existing | No | No | Segmentation | Surface Water |
| H. Xiao et al., 2019 | CNN | PyTorch | Existing | No | No | Classification | Surface Water |
| Mei et al., 2019 | CNN | PyTorch | Existing | No | No | Regression | Surface Water |
| Ling et al., 2019 | CNN | MATLAB | Acquired | No | No | Segmentation | Surface Water |
| Song et al., 2019 | CNN | Caffe | Existing | No | No | Classification | Surface Water |
| Hrnjica and Bonacci, 2019 | LSTM | - | Acquired | No | No | Sequence Prediction | Surface Water |
| Qi et al, 2019 | LSTM | - | Acquired | No | No | Sequence Prediction | Surface Water |
| Senlin Zhu et al., 2020 | LSTM | ANNdotNET, CNTK | Acquired | No | No | Sequence Prediction | Surface Water |
| Lee and Lee, 2018 | RNN, LSTM | - | Acquired | No | No | Regression | Water Quality |
| Hamshaw et al., 2018 | RBM | MATLAB | Acquired | Yes | Yes | Classification | Water Quality |
| P. Liu et al., 2019 | LSTM | Keras, TensorFlow | Acquired | No | No | Sequence Prediction | Water Quality |
| Yurtsever and Yurtsever, 2019 | CNN | Caffe | Collected | No | No | Classification | Water Quality |
| Li et al., 2019 | LSTM, GRU, ENN | - | Acquired | No | No | Sequence Prediction | Water Quality |
| Shin et al., 2019 | LSTM | - | Acquired | No | No | Sequence Prediction | Water Quality |
| Banerjee et al., 2019 | ANN | R/H2O | Collected | No | No | Regression | Water Quality |
| P. Wang et al., 2019 | LSTM | - | Acquired | No | No | Sequence Prediction | Water Quality |
| Yim et al., 2020 | ANN | MATLAB | Collected | No | No | Regression | Water Quality |
| Zou et al., 2020 | LSTM | Keras | Acquired | No | No | Sequence Prediction | Water Quality |
| Liang et al., 2020 | LSTM | Keras | Acquired | No | No | Sequence Prediction | Water Quality |
| Barzegar et al., 2020 | CNN, LSTM | - | Collected | No | No | Sequence Prediction | Water Quality |
| Yu et al., 2020 | LSTM | - | Acquired | No | No | Sequence Prediction | Water Quality |
| Duo Zhang et al., 2018a | LSTM | Keras, TensorFlow | Acquired | No | No | Sequence Prediction | Water Resources Management |
| Duo Zhang et al., 2018b | LSTM, GRU | Keras, TensorFlow | Acquired | No | No | Sequence Prediction | Water Resources Management |
| Duo Zhang et al., 2018c | LSTM, NARX, ENN | Keras, TensorFlow | Collected | Yes | Yes | Regression | Water Resources Management |
| Harrou et al., 2018 | DBN | - | Collected | No | No | Sequence Prediction | Water Resources Management |
| Shi and Xu, 2018 | AE | - | Collected | No | No | Regression | Water Resources Management |
| Zhou et al., 2019 | ANN, CNN | PyTorch | Existing | Yes | Yes | Classification | Water Resources Management |
| Fang et al., 2019 | CNN | - | Acquired | No | No | Classification | Water Resources Management |
| Sun et al., 2020 | ANN | - | Acquired | No | No | Regression | Water Resources Management |
| Karimi et al., 2019 | ANN, LSTM | MATLAB | Collected | No | No | Sequence Prediction | Water Resources Management |
| Xu et al., 2020 | LSTM | - | Acquired | No | No | Regression | Water Resources Management |
| Nam et al., 2020 | DQN | - | Acquired | No | No | Reinforcement Learning | Water Resources Management |
| Mamandipoor et al., 2020 | LSTM | Keras, TensorFlow | Acquired | No | No | Classification | Water Resources Management |
| Tang et al., 2018 | ANN | - | Acquired | No | No | Regression | Weather |
| Klampanos et al., 2018 | AE, CNN | - | Acquired | No | No | Unsupervised Learning | Weather |
| Scher and Messori, 2018 | CNN | Keras, TensorFlow | Acquired | No | No | Matrix Prediction | Weather |
| Ukkonen and Mäkelä, 2019 | ANN | Keras | Acquired | No | Yes | Classification | Weather |
| B. He et al., 2019 | LSTM | - | Acquired | No | No | Classification | Weather |
| Jeppesen et al., 2019 | CNN | Keras, TensorFlow | Existing | Yes | Yes | Segmentation | Weather |
| Chen et al., 2019 | GAN | PyTorch | Acquired | No | No | Matrix Prediction | Weather |
| Wieland et al., 2019 | CNN | Keras, TensorFlow | Existing | No | Yes | Segmentation | Weather |
| Weyn et al., 2019 | CNN, LSTM | Keras, TensorFlow | Acquired | No | No | Matrix Prediction | Weather |
| Wei and Cheng, 2019 | ANN | Weka | Acquired | No | No | Sequence Prediction | Weather |
| Z. Zhang et al., 2019 | CNN | - | Acquired | No | Yes | Classification | Weather |
| Kim et al., 2019 | ANN | - | Acquired | No | No | Regression | Weather |
| Pan et al., 2019 | CNN | - | Acquired | No | Yes | Matrix Prediction | Weather |
| Tran and Song, 2019 | LSTM, GRU, CNN | TensorFlow | Existing | No | Yes | Matrix Prediction | Weather |
| Poornima and Pushpalatha, 2019 | LSTM, ELM | Keras, TensorFlow | Acquired | No | Yes | Sequence Prediction | Weather |
| Chai et al., 2019 | CNN | - | Acquired | No | Yes | Segmentation | Weather |
| Wu et al., 2020 | LSTM, CNN | - | Acquired | No | Yes | Matrix Prediction | Weather |
| Zhang et al., 2020 | LSTM | - | Acquired | No | No | Regression | Weather |
| Su et al., 2020 | LSTM, CNN | - | Acquired | No | No | Sequence Prediction | Weather |
| Chen et al., 2020 | CNN, LSTM | PyTorch | Existing | No | Yes | Matrix Prediction | Weather |
| Weber et al., 2020 | CNN | TensorFlow | Acquired | Yes | Yes | Matrix Prediction | Weather |
| Yan et al., 2020 | CNN | TensorFlow | Acquired | No | Yes | Sequence Prediction | Weather |
| Q. Wang et al., 2020 | RNN | - | Acquired | No | No | Matrix Prediction | Weather |

## Results

In this section we provide brief summaries of papers presented in the previous section (Table 1.) Papers are grouped and interpreted by their use case.

## Streamflow and Flood

Runoff prediction and flood forecasting are major tasks in rainfall-runoff modeling. Toward this end, many researchers have applied cutting-edge deep learning architectures to the runoff prediction and flood forecasting tasks. **Since rainfall and runoff are both time series data, the common networks for the streamflow prediction and flood forecast are RNN, LSTM, NAR,** **and ENN.** Kratzert et al. (2018) applied an LSTM model on daily runoff prediction for the first time, considering the meteorological observations with results better than a well-established physical model SAC-SMA+Snow-17. In 2019, Kratzert et al. (2019) further applied the LSTM model to 531 watersheds in U.S. with k-fold cross validation and it shows that LSTM can be applied on ungaged watersheds with better results than physical models such as calibrated SAC-SMA and the National Water Model. Other researchers applied recurrent neural networks to the runoff forecast and compared their outputs to other machine learning models. Damavandi et al. (2019) proposed an LSTM model on a Texas watershed predicting the next day's daily streamflow using climate data and the current day's streamflow. Their results show that LSTM performs better than physical model CaMa-Flood. Di Zhang et al. (2018) applied an LSTM model on monthly reservoir inflow and outflow predictions in hourly, daily, and monthly basis with better results than SVM and BPNN. Kumar et al. (2019) applied RNN and LSTM models for the monthly rainfall prediction in India and shows LSTM provides better results. Qin et al. (2019) applied the LSTM on the streamflow prediction and compared it with the Autoregression model. S. Yang et al. (2019) applied an LSTM model on the daily reservoir overflow prediction. NAR with external input and LSTM are used for three reservoirs with geographic information, daily precipitation, air temperature, wind speeds, relative humidity, and sunshine duration. Wan et al. (2019) also successfully applied Elman Neural Network to a real-time 3-hour ahead flood forecasting.

**Since hyper-parameter optimization is a problem in deep learning, some studies focused on applying additional optimization algorithms on deep learning models.** Yuan et al. (2018) proposed two models that use the ant lion optimizer (ALO) and particle swarm optimization (PSO) to optimize the parameters and hidden layers of an LSTM model, which are named LSTM-ALO and LSTM-PSO. Their results show that the LSTM-PSO outperformed the LSTM and LSTM-ALO. S. Yang et al. (2019) proposed a Genetic Algorithm based NAXR (GA-NAXR), which outperforms the NAXR and LSTM. Ni et al. (2019) proposed two LSTM based models, which are wavelet-LSTM (WLSTM) and CNN+LSTM (CLSTM) for the rainfall and streamflow forecasting, and both models have better results than LSTM. Kabir et al. (2020) proposed a wavelet-ANN to make the hourly streamflow predictions. Results show that wavelet-ANN can make acceptable predictions for at most 6 hours ahead, which outperforms ANN, DBN, and SVR.

**Some studies focused on the data pretreatment such as the decomposition of rainfall and runoff prior to deep learning models.** He et al. (2019) proposed a DNN model for daily runoff prediction where the inputs are the runoff series that were decomposed into multiple intrinsic mode functions (IMFs) with variational mode decomposition (VMD). Hu et al. (2019) proposed an LSTM model for flood forecast with the preprocessing of proper orthogonal decomposition (POD) and singular value decomposition (SVD) prior to LSTM. Zuo et al. (2020) proposed three LSTM models with different pretreatments, which are variational mode decomposition (VMD), ensemble empirical mode decomposition (EEMD), and discrete wavelet transform (DWT) for daily streamflow up to 7 days ahead. Zhu et al. (2020) proposed a probabilistic LSTM model coupled with the Gaussian process (GP) to deal with the probabilistic daily streamflow forecasting. These studies show that the pretreatment of input may help to improve the deep learning model accuracy.

**Other studies focused on constructing more complex deep learning architectures such as autoencoder, encoder-decoder, and customized layer based on the LSTM.** Bai et al. (2019) proposed an LSTM model with stack autoencoder (SAE) to predict daily discharge values based on one-week discharge values, and the results of SAE-LSTM outperform LSTM alone. For multiple time-step flood forecast tasks, an encoder-decoder LSTM is proposed for the runoff prediction by Kao et al. (2020) and Xiang et al. (2020). Kao et al. (2020) proposed an encoder-decoder LSTM model that can be used on multi-timestep output predictions for up to 6 hours. Xiang et al. (2020) proposed an encoder-decoder LSTM model that can be used to predict for up to 24 hours ahead. Both studies showed the encoder-decoder LSTM is better than LSTM. In particular, Kratzert et al. (2019) proposed the Entity-Aware-LSTM network, which designed a specific layer for the rainfall-runoff modeling based on LSTM. This network allows for learning catchment similarities as a feature layer, and data from multiple watersheds can be used to provide runoff for a watershed. These studies applied some high-level designs to the LSTM network and can perform much better than normal LSTM on the long-term or multiple watersheds.

**Several studies coupled physical models with deep learning networks.** T. Yang et al. (2019) proposed a model using LSTM to improve the performance of flood simulations of a physical model. The watershed-averaged daily precipitation, wind, temperature and model-simulated discharge from GHMs+CaMa Flood model in 1971-2020 were used to model the discharge. **This indicates the deep learning models can be used to improve the streamflow forecast accuracy of physical models.** Nguyen and Bae (2020) proposed an LSTM network using the quantitative precipitation forecasts (QPFs) of the McGill Algorithm for Precipitation nowcasting by Lagrangian Extrapolation (MAPLE) system to reproduce three-hour mean areal precipitation (MAP) forecasts. Corrected MAPs are used as input to a coupled 1D/2D urban inundation model to predict water levels and relevant inundation areas (1D conduit network model and 2D overland flow model). In this model, LSTM takes a forecast from MAPLE and then reproduced MAPs are used by coupled 1D/2D urban inundation model. Worland et al. (2019) proposed a DNN model to predict the flow-duration curves using USGS streamflow data by constructing 15 output values representing 15 quantiles of the curve. **These indicate the deep learning models can be used as surrogate models of physical models or curves.**

**Some studies focused on flood susceptibility and flood prediction capabilities.** Bui et al. (2020) applied a DNN network with 11 variables such as DEM, aspect, slope, etc. to predict the flood susceptibility, a value between 0 to 1, for an area. Wang et al. (2020) proposed a CNN network with 13 flood triggering factors in the study area to map the flood susceptibility for the study area. In this project, each pixel may have different flood susceptibility values. Sankaranarayanan et al. (2019) applied a DNN network to classify if flooding is possible in an area for the month given total precipitation and temperature. This study simplified the project into a binary prediction task of if there will be a flood event or not. The same simplification is done by Tien et al. (2020), who applied a DNN network to classify if a region is flash flood prone or not using the elevation, slope, aspect, curvature, stream density, NDVI, soil type, lithology and rainfall data as input.

**With CNNs, some innovative flood monitor and forecast projects are now possible.** Wang et al. (2019) applied a CNN model that predicted the real-time hourly water levels for warning systems during typhoons using satellite images. Bhola et al. (2019) proposed a model that estimates the water level from CCTV camera images using a CNN model. The model performed the edge detection to find the water body in the photo, and the water level is then calculated from some physical measurements. Moy et al. (2019) proposed a model that calculated the flooded-area and the surface observed flooding index (SOFI) with CCTV cameras using CNN. U-net CNN is used to segment the stream shape from a CCTV camera and another CNN model is used to calculate the flooded areas in pixels from the segmented photo.

**Subsurface and Groundwater**

There are several different types of studies in subsurface and groundwater. Different deep learning models can be used in each of different types of studies. **One type of groundwater study is the estimation of water table level or flow rate.** With the groundwater monitoring wells data, this is a 1-D regression task similar to the surface water predictions using recurrent models. Jeong and Park (2019) applied the NARX, LSTM and GRU on the water table level estimations with observed data from monitoring wells. It is found that the estimations from the NARX and LSTM models are superior to those of the other models in terms of prediction accuracy. Jeong et al. (2020) further applied LSTM with multiple loss functions, which shows that the proposed LSTM model with cost function of MSE with Whittaker smoother, least trimmed squares, and asymmetric weighting has the best performance on groundwater level prediction with data corrupted by outliers and noise. For groundwater flow prediction, Wang et al. (2020) proposed a theory-guided deep neural network which not only makes the predictions of groundwater flow, but also estimates the parameters of the partial differential equation (PDE) as well as the initial condition and boundary condition of the PDE. By constructing the loss function, the theory-guided neural network can provide the prediction results with reasonable parameters of the physical model PDE.

**More groundwater studies take a cross section in time-series and focus on a 2-D map.** Examples include the groundwater water balance map (Sun et al., 2019), hydraulic conductivity field map (Zhou et al., 2020), pressure and $CO_2$ saturation field map (Mo et al., 2019a). CNN models can be used in these studies. Sun et al. (2019) proposed a CNN model using the physical model NOAH simulation results as input to predict the groundwater water balance. Thus, the CNN model is used to correct the physical model results by learning the spatial and temporal patterns of residuals between GRACE observations and NOAH simulations. Results show that the CNN can significantly improve the physical model simulation accuracy. Zhou et al. (2020) proposed a CNN of eight layers to learn a map between stochastic conductivity field and longitudinal macro-dispersivity based on synthetic 2-D conductivity fields. The estimations are in acceptable accuracy with moderate heterogeneity. Mo et al. (2019a) proposed an Encoder-Decoder CNN to approximate the pressure and $CO_2$ saturation field map in different time steps as a surrogate model. **Furthermore, some studies in groundwater concern 3D mapping.** An example is the flow rate estimation in a 3D rock. Santos et al. (2020) proposed a 3D CNN to predict the estimated state solution of Navier-Stokes equation for laminar flow, which is the flow rate, with 3D rock images. This 3D CNN is a surrogate model of the Navier-Stokes equation, and takes less than a second.

**In groundwater studies, one of the other deep learning applications is inversion, such as to identify the contaminate source in groundwater.** Some complex traditional algorithms can be

used to solve this problem. However, it will be not efficient when the data is in high dimensions. Mo et al. (2019b) developed a deep autoregressive neural network which was used as the surrogate model of this high dimensional inverse problem and provides more accurate inversion results and predictive uncertainty estimations than physical models. Laloy et al. (2018) proposed another approach for the inversion by using GAN. With the GAN models, the inversion rapidly explores the posterior model distribution of the 2-D steady state flow and recovers model realizations that fit the data close to the true model results.

Finally, Sun et al. (2020) investigate three learning-based models, namely DNN, Multiple Linear Regression (MLR), and SARIMAX, to find missing monthly data in total water from GRACE (Gravity Recovery and Climate Experiment) Data (Tapley et al. 2004). Based on the results, the performance of DNN is slightly better than SARIMAX, significantly better than MLP in most of the basins. However, three learning-based models are reliable for the reconstruction of GRACE data in areas with humid and no/low human interventions. Dynamic multiphase flow in heterogeneous media is a hard problem in groundwater studies. These studies show the deep learning models can work as surrogate models by improving computational efficiency in groundwater dynamic predictions and inversions.

**Land and Soil**

The segmentation and classification of land use and land cover are important for water and soil resources management. **Many studies have applied the deep learning networks to create a more accurate land cover map from satellite or radar imagery.** Abdi et al., 2018 applied SAE and CNN on land cover classification of urban remote sensing data. Tests on 9 datasets show that SAE and CNN are better than machine learning models like logistic regression, Naive Bayes, KNN and SVM. Cao et al. (2018) applied ELM and CNNs to classify land cover categories on satellite imagery and the results show the combined CNN–ELM method has the highest accuracy. Shen et al. (2019) applied a DNN network to predict the land drought index using the precipitation data, meteorological drought index data and soil relative moisture data. Bhosle et al. (2019) applied a CNN model on the land cover land use classification on Indian Pines dataset. Some studies applied more complex coupled models including encoder-decoder, autoencoder, 3D networks, and coupled machine learning models to achieve a higher model accuracy. Zhang et al. (2019) applied 3D-CNN and 3D-DenseNet models on the land cover land use classification on Indian Pines and Pavia University dataset. All these studies showed a high accuracy of the land cover identification task. Kounghoon et al. (2020) applied the autoencoders prior to the random forest, SVM and other machine learning models to the landslide susceptibility prediction. Results show that random forest with an autoencoder gives more accurate results than other machine learning methods. In the study of wetland type identification by Meng et al. (2019), results show that the ensemble model SVM-CNN performs better than CNN and SVM. O'Neil et al. (2020) applied an Encoder-Decoder CNN model to classify the wetland types using LiDAR radar dataset and NDVI index data. This study used the LiDAR DEM and remote sensing images to generate physically informed input including Slope, NDVI, DTW, and TWI. In particular, Kroupi et al. (2019) applied a CNN based model to land cover classification that trained on the European satellite dataset - EuroSAT and tested in a region outside Europe, and still provides promising results despite differences in tested and trained regions. These results indicate the CNN models have a high model accuracy as well as high robustness in the land use land type classification task.

**In addition to applying some known networks, some researchers developed objective-based CNN rather than traditional pixel-based CNN to better identify the land use and land types.** Zhang et al. (2018) proposed an object-based CNN to label very fine spatial resolution (VFSR) remotely sensed images to do object-wise segmentation rather than pixel-wise segmentation in urban areas. Fu et al. (2018) proposed a blocks-based object-based CNN for classification of land use and land cover types and achieved end-to-end classification. This model works well on irregular segmentation objects which is a common in land use classification. Jiang (2018) proposed an object-based CNN with an autoencoder for extracting high level features automatically. Results show the AE-object-CNN is better than three manual design feature systems. Jin et al. (2019) proposed an object-oriented CNN which used a typical rule set of feature objects to construct the object-oriented segmentation results before using a CNN to make the classification. These studies show object-based studies have better performance than simple CNN.

**Some studies focus on more specific tasks based on different study purposes such as the identification of agricultural fields, impervious surfaces, wetland types, water body types, and crop types.** Persello et al. (2019) applied two CNN models, SegNet (Badrinarayanan et al., 2017) and VGG16 (Simonyan and Zisserman 2014), for the segmentation and classification of agricultural fields. Sun et al. (2019) applied a CNN model to classify if the land is impervious surfaces, vegetation or bare soil from satellite imagery or both the satellite and LIDAR data. Results show that the model with LIDAR data provides better results than the model with satellite imagery only. Meng et al. (2019) applied to identify the wetland types in one lake using the Chinese remote sensing imagery GF-2. Bhosle et al. (2019) applied a CNN model to identify crop types on EO-1 Hyperion sensor hyperspectral imagery. Yang et al. (2020) applied CNNs to identify the water body types from remote sensing images. Mask R-CNN is used to segment the water body, and the ResNets (He et al., 2015) including ResNet-50 and ResNet-101 are used to identify the water body types. Results show a high accuracy on regular-shaped water bodies.

**Similar networks can be used to generate DEMs, which is another type of land use study.** Jiang et al. (2019) applied a CNN model to predict the paleo-valley DEM using the original DEM data and electrical conductivity. The electrical conductivity data are collected from the field study, and the CNN models used in this study distinguish the valley and non-valley pixels, which can find the spatial connectivity of the paleo-valley. The CNN can efficiently constrain three-dimensional paleo-valley geometry DEM.

**Snow cover, a special study in land cover studies, can be measured by deep learning models in different approaches.** Kopp et al. (2019) proposed a model to predict the snow depth using OpenCV and Mask R-CNN on the surveillance camera photos. The Mask R-CNN is used to segment the detectable measuring rod, and then the OpenCV library can be used to identify the snow depth by measuring how much of the measuring rod gets covered by snow. Wang et al. (2020) applied a DBN and CNN to estimate snow depth from the geographical data. DBN takes multiple inputs including latitude, longitude, elevation, forest cover fraction, time, passive microwave horizontal and vertical polarization brightness temperatures of 19 and 37 GHz. Results show that the DBN outperforms CNN in this study.

**Additional land and soil related studies include the land surface temperature, soil salinity, vegetation dynamics over time, and these can be done with time-series related models such as LSTM.** Zhang et al. (2018) applied the ensemble composition-based LSTM models on the daily land surface temperature simulation. The original daily land surface temperature data series were decomposed into many Intrinsic Mode Functions (IMFs) and a single residue item. And the Partial Autocorrelation Function (PACF) is used to obtain the number of input data sample points for LSTM models. Zeng et al. (2018) applied Partial Least Square Regression (PLSR), SVM and DNN for predicting soil salinity from images. Surprisingly, DNN performs worse than PLSR for this task. Reddy et al. (2018) applied a LSTM model to predict the vegetation dynamics using NDVI images from 2000 to 2016 in a 7-day gap. This study shows that the single feature NDVI can be used to provide accurate vegetation prediction over time.

**Surface Water**

**The prediction of water level is crucial for water resources management and protecting the natural environment. Many studies have applied deep learning methods, such as LSTM and ANN, to forecast the water level from one day to one year.** Hrnjica and Bonacci (2019) investigate two different ANNs, namely LSTM and Feed Forward Neural Network (FFNN), to forecast the lake water level. Monthly measurements for the last 38 years of Vrana Lake, Croatia is utilized for training the models in order to predict 6 or 12 months ahead of the lake water level. The set of sequences with different lengths created from the obtained data is used in the networks, instead of using classical lagged data. The results of LSTM and FFNN are compared with classical time forecasting methods and ANN. According to the results, the performance of LSTM is the best among the models in all scenarios, while FFNN provides better accuracy than the compared methods in both 6 and 12 months prediction. Senlin Zhu et al. (2020) also investigate LSTM and FFNN in their work with data from a different region. However, the models in this paper are designed to predict one month ahead of the lake water level for 69 temperate lakes in Poland. Their results indicate that LSTM and FFNN perform similarly most of the time, unlike in the previous paper. The reasons for this situation can be the differences between the datasets, prediction intervals, or model designs.

In addition to lake water level prediction, Yongqi Liu et al. (2019) developed a Bayesian Neural Network, which is based on ANN with posterior inference, to forecast the water level in a reservoir to derive operation rules. According to the paper, the current reservoir status and future inflows are the primary factors that affect operational decisions. Also, the influence of inflow uncertainties on reservoir operations is more than model parameter uncertainty. Their findings show the impact of the input data alongside the promising results of Bayesian NN. Qi et al. (2019) forecast daily reservoir inflow by ensembling the different results from the LSTM models with different decomposed inflow data as inputs for more accurate assumptions. Zhang et al. (2018) propose an LSTM model to predict water table depth in agricultural areas. The LSTM model takes monthly water diversion, evaporation, precipitation, temperature, and time as inputs for prediction of the water table depth. The results of the LSTM model are compared with the FFNN model, and the LSTM model performs better than the FFNN model. Also, it is highlighted that the dropout method increases the LSTM model's accuracy in this task.

**Alongside the water level or flow, the prediction of water temperature has received much attention in the scholarship.** C. Xiao et al. (2019) propose a convolutional long short-term

memory (ConvLSTM) model to predict the sea surface temperature (SST). In the paper, 36 years of satellite-derived SST data are used. Based on the results, the ConvLSTM model produces more accurate results than the linear support vector regression model and two different LSTM models for short and mid-term temperature prediction. Read et al. (2019) aim to predict lake water temperatures based on depth. A hybrid LSTM and theory-based feedbacks (model penalties) model was developed. According to the results, the hybrid model produces the best results among the tested methods in the paper. The results from the paper can be seen as an example of improving predictions by integrating scientific rules with DL methods.

**In addition to the aforementioned subtopics in surface water, various tasks are also investigated by scholars such as segmentation, change detection, or super-resolution.** Despite extensive usage of LSTM or ANN in previously aforementioned papers, CNN models are generally used in these tasks. Yansong Liu et al. (2019) label objects in aerial images as water, tree, etc. with the help of a fully convolutional neural network (FCN) model and multinomial logistic regression. FCN takes the aerial image and returns a probability. At the same time, LiDAR data passes into multinomial logistic regression and returns another probability. These two probabilities are combined with higher-order conditional random field formulation to produce a final probability. Based on the final probability, objects are labeled with the corresponding group. Mei et al. (2019) develop a CNN based model to measure the sea ice thickness in Antarctica from LiDAR data. The input is a windowed LiDAR scan (snow freeboard), and the mean ice thickness is the output of the model. In addition to the LiDAR scan, the paper investigates the effects of different inputs such as surface roughness and snow depth on the task. Ling et al. (2019) use the CNN model in order to generate a finer resolution of the image to measure the wetted river width. Then, the output of CNN is used to measure the width of the river. Song et al. (2019) aim to detect the change in surface water in remote sensing images. An FCN model is proposed and used for the same regions at different times. The FCN model returns the surface water regions for each time. The outputs for the same regions are compared, and a change map is created to show the difference. Beyond these publications, H. Xiao et al. (2019) introduce a dataset for the classification of ice crystals. The dataset contains 7282 images in 10 different categories. The performance of numerous pre-trained models, such as AlexNet (Krizhevsky et al. 2012) and VGGNet(s) (Simonyan and Zisserman 2014), are also provided in the paper.

**Water Quality**

**Water quality monitoring and prediction are vital operations for many fields, such as water resources management and water treatment.** Water quality and safety depend on numerous parameters with complex biological, chemical, and physical interactions. As such, deterministic water quality models are a realistic option in only the simplest and idealized scenarios. However, data-driven models are increasingly being used in a variety of water quality applications. One such application area is predicting surface water quality. Li et al. (2019) propose an ensemble approach that combines three RNN models with Dempster/Shafer (D-S) evidence theory (Shafer 1976) to predict the quality of water. The results of three RNN models, namely LSTM, GRU, and Elman Neural Network, are combined by D-S evidence theory for the final output. The combined model predicts at most 50 hours in advance, and the results show that the model accuracy reduces significantly over 25 hours. Liu et al. (2019) use an LSTM model to forecast drinking water quality for up to 6 months. Zou et al. (2020) develop an ensemble approach to

predict water quality data, such as pH, DO, CODMn, and NH3-N. The approach based on using three LSTM models that different size interval data feed each of them and the final prediction is a combination of the results of three LSTM models. Banerjee et al. (2019) choose the indicators, namely dissolved oxygen and zooplankton abundance, to reflect the water quality level of a reservoir.

An ANN model is proposed to model the selected indicators in order to represent the water quality level. Yu et al. (2020) combine the LSTM model with Wavelet Mean Fusion and Wavelet Domain Threshold Denoising to simulate the change of chlorophyll-a concentration in Dianchi Lake, China and use 15 water quality parameters as inputs, such as pH and DO. Liang et al. (2020) also work with the prediction of the chlorophyll-a concentration level. Fabricated data are created by the environmental fluid dynamics code (EFDC) to train an LSTM model. The LSTM model can forecast the chlorophyll-a concentration level for up to one month. Chlorophyll-a, water temperature, and total phosphorus are identified as critical inputs that affect the performance of the LSTM model. Barzegar et al. (2020) investigate multiple models to predict the level of DO and chlorophyll-a in Small Prespa Lake, Greece. Three different NN models, namely CNN, LSTM, and CNN-LSTM, were developed to forecast the DO and chlorophyll-a concentrations and use pH, oxidation-reduction potential (ORP), water temperature, and electrical conductivity (EC) as inputs for the models. In addition to ANN models, SVM and decision tree models are used for the performance comparison. According to results, the hybrid CNN-LSTM model provides the best accuracy to predict both DO and chlorophyll-a.

Yim et al. (2020) develop a stacked autoencoder-deep neural network (SAE-DNN) to predict phycocyanin (PC) concentrations in inland waters from in-situ hyperspectral data. The proposed architecture's ability for the prediction from airborne hyperspectral imagery is examined. Shin et al. (2019) introduce an LSTM model to forecast the occurrence of harmful algal blooms in the South Sea of Korea. Sea surface temperature and photosynthetically available radiation are extracted from satellite data to be used as inputs for the LSTM model in order to minimize the damage. Lee and Lee (2018) aim to predict the occurrence and number of harmful algal blooms with an LSTM model by providing weekly water quality and quantity data. The LSTM model's performance is compared with RNN and MLP models, and the LSTM provides better results among all the investigated methods based on the results. Hamshaw et al. (2018) use the Restricted Boltzmann Machine (RBM) to classify the sediment-discharge curve in 14 categories from the 2D image of the suspended-sediment discharge plots from 600+ storm events. Finally, an LSTM based system is proposed to identify the characteristics of the water pollutants and trace its sources in the work of Wang et al. (2019). In the system, a water quality cross-validation map is generated to identify pollutants and, based on defined rules, track the pollutants to common industries.

**Water Resources Management**

**Urban water systems are essential to modern cities. Efficient operation of water treatment plants, wastewater treatment plants, and conveyance networks require accurate modelling of these interconnected systems.** Duo Zhang et al. (2018b) investigate the multiple models to simulate and predict the water level of Combined Sewer Overflow (CSO) structure. In the study, the collected data from IoT is used separately with four different neural networks, namely MLP,

ANN with Wavelet, LSTM, and GRU, to compare the networks with each other. According to the results, LSTM and GRU have good performances with respect to others, but GRU has a quicker learning curve, fewer parameters, and simpler architecture. Despite these advantages, the accuracy of GRU is slightly lower than the LSTM. Duo Zhang et al. (2018a) predict the next hour's wastewater flow to avoid sewer overflow using LSTM with the traditional hydraulic model on the sewer system. This one-hour prediction is tested in several scenarios, which are: 1-time step prediction at the 1-hr sampling frequency, 2-time steps prediction at the 30-min sampling frequency, 4-time steps prediction at 15-min sampling frequency, and 6-time steps prediction at 10-min sampling frequency. The performance of the LSTM based model is compared with Support Vector Regression (SVR) and Feed Forward Neural Network (FFNN), and the LSTM based model has the best accuracy in all scenarios. Duo Zhang et al. (2018c) also predict the next hour's wastewater inflow for the wastewater treatment plant. The paper aims to identify which parts of the sewer system have more free space and take action based on the outcome. LSTM, NARX, Elman Neural Network are compared, and the LSTM model provides better results than other methods based on the results. Karimi et al. (2019) propose an LSTM model to forecast flow in sanitary sewer systems. It is claimed that accepting the groundwater as an additional input for the LSTM model increases the overall accuracy of the task.

**In addition to the prediction of the wastewater level or flow, some studies aim to detect conveyance network conditions.** Xu et al. (2020) aim to detect abnormal working conditions in the water supply network. Besides the detection of abnormal conditions, pressure in the water supply network is predicted. An LSTM model is developed to achieve these goals. The performance of the LSTM model outperforms the traditional prediction models, such as SVM. Zhou et al. (2019) use an ANN model to identify burst locations in a water distribution network. The model takes the pressure data as input and returns one or several possible pipes, which can be the location of the burst. Fang et al. (2019) focus on detecting multiple leakage points in a water distribution network with a CNN based model. The model accepts the pressure measurements of the distributed water system as input and returns the possible locations of leakage points.

**Various studies explore the strength of deep learning powered modeling in water and wastewater treatment procedures.** Shi and Xu (2018) develop a Stacked Denoising AutoEncoders (SDAE) to predict a two-stage biofilm system's performance. Nam et al. (2020) propose a DQN-based system to operate membrane bioreactor (MBR) more efficiently. It aims to maximize the system's energy efficiency while meeting stringent discharge qualities. GoogLeNet (Szegedy et al. 2015) architecture is used to identify and classify microbeads in urban wastewater into five categories based on microscopic images in the work of Yurtsever and Yurtsever (2019). Harrou et al. (2018) provide a case study using a DBM-SVM model to identify abnormal signals from the water features such as pH, conductivity, etc. in wastewater treatment plants. The results show that it is possible to detect the abnormal conditions in order to alert the system early based on the outcome of the DBM-SVM model. Mamandipoor et al. (2020) develop an LSTM model to monitor a wastewater treatment plant for detecting faults during the oxidation and nitrification processes.

**Weather**

**Rainfall forecasting is one of the significant tasks in the domain of meteorology. Several techniques have been proposed to forecast rainfall with the help of statistics, machine learning, and deep learning.** Zhang et al. (2020) introduce an ensemble approach to forecast rainfall. In the first step, eight major meteorological factors are selected via correlation analysis between control forecast meteorological factors and real-time rainfall. Then, samples are divided into four categories by K-means clustering. The LSTM based model is fed by each cluster, and outputs are combined to reach the final prediction. Weber et al. (2020) develop a CNN-based surrogate model for one of the global climate model CanESM2 (Arora et al. 2011). The CNN model is fed by 97 years of monthly precipitation output from CanESM2, and the model preserves its performance even when the forecast length is expanded to 120 months. According to the paper, the accuracy of the model can be increased by deeper networks. Poornima and Pushpalatha (2019) investigate an LSTM model to forecast the rainfall with the help of 34 years of rainfall data. In the paper, the LSTM model results are compared with multiple methods, such as Holt-Winters and ARIMA. Tang et al. (2018) introduce a DNN model to predict rain and snow rates at high altitudes. In the paper, passive microwave, infrared and environmental data are trained to the reference precipitation data sets, which are obtained by two space-borne radars for the estimations. The results of the DNN model are compared with many methods, such as the Goddard Profiling Algorithm (GPROF). The experiment results show that the DNN model is capable of predicting snow and rain rate more accurately than other tested methods at high altitudes.

**Some studies specifically focus on the improvement of quantitive precipitation estimation accuracy, alongside precipitation nowcasting.** Since the prediction of precipitation generally is a time series problem, the usage of LSTM architecture is common on this task. Wu et al. (2020) design a fusion model, which is a combination of CNN and LSTM, to improve quantitative precipitation estimation accuracy. The proposed model uses satellite data, rain gauge data, and thermal infrared images. The CNN part of the model extracts the spatial characteristics of the satellite, rain gauge, and thermal infrared data, where The LSTM part of the model handles the time dependencies of the provided data. The performance of the CNN-LSTM model is better than the comparative models, such as CNN, LSTM, and MLP. Yan et al. (2020) use a CNN model to forecast short-term precipitation with the help of radar reflectance images for a local area in China. As a dataset, the radar reflection images and the corresponding precipitation values for one hour are collected. The model takes the images as inputs and returns the forecast value for one-hour precipitation. The CNN model contains residual links between the layers, which increase the efficiency of the model. Chen et al. (2020) focus on precipitation nowcasting using a ConvLSTM model. The model accepts the radar echo data in order to forecast 30 or 60 minutes of precipitation value. According to results, using the customized multisigmoid loss function and group normalization provides better performance than ConvLSTM with classical loss functions, such as cross-entropy loss function, and conventional extrapolation methods. Statistical downscaling methods often provide more accurate precipitation estimation than using raw precipitation values in the models.

**Deep learning methods can be used as statistical downscaling methods to improve the accuracy of the tasks.** Pan et al. (2019) propose a CNN model as a statistical downscaling method (SD) for daily precipitation prediction. The method is tested with 14 geogrid points in the U.S., and SD results from the CNN model outperform other tested methods, including linear regression, nearest neighbor, random forest, and DNN. Wang, Q. et al. (2020) develop an RNN model to perform statistical downscaling on temperature and precipitation in order to improve the accuracy of hydrological models. The RNN model provides better accuracy than the compared methods, such as ANN, for the extreme temperature and precipitation downscaling based on the evaluation of downscaled data on the Soil and Water Assessment Tool (SWAT) model.

**In meteorology and remote sensing, cloud or cloud shadow detection has received attention in the recent literature.** Because U-net (Ronneberger 2015), a CNN model for biomedical image segmentation, and Segnet (Badrinarayanan 2017), a convolutional encoder-decoder architecture for image segmentation, provide successful results in their domain, many researchers use those models as a base model for their works. Jeppesen et al. (2019) use the U-net to detect clouds in satellite imagery. The model is trained and tested with Landsat 8 Biome and SPARCS datasets. Wieland et al. (2019) aim to segment cloud, shadow, snow/ice, water, and land in multi-spectral satellite images. The U-net based model is proposed and trained with Landsat datasets to segment images into five categories. According to results, contrast and brightness augmentations of the training data improve the segmentation accuracy, alongside adding shortwave-infrared bands. Z. Zhang et al. (2019) applied the U-net based model on the red, green, blue, and infrared waveband images from the Landsat-8 dataset for cloud detection. LeGall-5/3 wavelet transform is used on the dataset to accelerate the model and make it feasible to implement on-board on satellite. Chai et al. (2019) propose a CNN model based on Segnet to detect clouds and cloud shadow in Landsat imagery.

**Thunderstorms and typhoons are one of the extreme natural disasters that can cause massive damages. Some studies focus on the prediction of those hazardous natural events to take early actions to minimize the damage.** Ukkonen and Mäkelä (2019) aim to predict the occurrence of thunderstorms from parameters related to instability, inhibition, and moisture mainly. A DNN model is trained with lightning data and a high-resolution global reanalysis. Various regions, such as Sri Lanka and Europe, are used as test areas for the model. Many valuable findings are provided related to the correlation between thunderstorm occurrence and parameters specific to regions. Kim et al. (2019) aim to find similarities between the typhoon and a typhoon from the past for helping to mitigate the effect of the typhoon. In the study, a DNN model is used to encode the typhoon event by typhoon parameters, such as route, pressure, and moving speed. The model returns the typhoon events' similarity to the historical ones, which provide insights for officials to take early action. During the work, a database is created for 189 typhoons that occurred between 1950 and 2017. Wei and Cheng (2020) aim to predict wind speed and wave height of a typhoon with the help of an RNN model, namely TSWP. The TSWP model predicts wind speed first, followed by wind height for 1-6 hours in the future. The results of the TSWP model outperform the comparative methods, such as MLP, DNN, and Logistic Regression.

**In addition to the aforementioned studies, various works have been published for different meteorological tasks.** Su et al. (2020) use the pyramid delaminating technique to generate the global optical flow field. A ConvLSTM model takes the RGB image and generated flow field in order to improve the forecast accuracy of echo position and intensity. The generation,

dissipation, and merging of convective cells were also better identified in comparison to other classical methods. Weyn et al. (2019) investigate to forecast weather at 500-hPa geopotential height by a CNN model. The paper aims to use historical gridded reanalysis data in the model without explicit knowledge about the physical process. The CNN model produces promising results for capturing the climatology and annual variability of 500 - hPa heights and predicts realistic atmospheric states for 14 days. However, the CNN model still could not perform as good as an operational weather model. B. He et al. (2019) propose an LSTM model to classify periods as rainy or dry by microwave links. Scher and Messori (2018) aim to predict the uncertainty of a forecast based on past forecasts and their occurrence rate with a CNN model. The CNN model takes the atmospheric fields at three different heights as inputs and returns the scalar value representing the predictability (or the uncertainty) of precipitation forecast. Klampanos et al. (2018) study the relationship between nuclear events and weather. A model with Autoencoder and CNN is used for rapid source estimation during radiological releases. The model clusters weather events first and looks over their correlations with nuclear events. Chen et al. (2019) design a GAN model based on SRGAN (Ledig 2017), which is a GAN architecture for single image super-resolution, to improve the resolution of radar echo images of weather radar systems in order to increase the accuracy of the tasks that accepts echo images as inputs. Tran and Song (2019) design a ConvRNN model for the radar echo extrapolation task. The model uses multiple (five) satellite images and predicts 10 steps ahead at the pixel level. During the training process, Structural Similarity (SSIM) and multi-scale SSIM are used to obtain better results.

**Unclassified Studies**

Some literature included in this review did not fit within any of the defined categories in this review. Summaries of these papers are provided in this section. Many of the papers in this section apply deep learning methods to ocean processes. In coastal hydraulics Kang (2019) used an improved CNN on images to classify and monitor waves. Also in coastal hydraulics, Kim et al. (2020) used a number of deep neural networks with coastal video imagery to develop a framework to track nearshore waves. Kylili et al (2019) use a CNN to identify floating plastic debris in ocean images. Further, Wang et al. (2019) used a CNN architecture with a satellite radar product to classify ocean surface roughness into ten geophysical phenomena. Li et al. (2018) also used a CNN to classify hurricane damage from post-event aerial imagery.

The remaining three papers in this section apply deep learning in a variety of disciplines. Haklidir and Haklidir (2019) use a DNN to predict the temperature of geothermal springs and wells given hydrogeochemical data, including chemical concentrations. Amirkolaee and Arefi (2109) implemented a very deep CNN structure to estimate a digital elevation model from single airborne or spaceborne images. Finally, Rohmat et al. (2019) developed and embedded a DNN into a GIS-based basin-scale decision support system to assess the impacts of best management practices (BMP.) Based on geographic information and potential BMP, the DNN returns multiple outputs which describe net flow separation between groundwater recharge, groundwater return flow, and overland flow. This project demonstrates the potential of deep learning to integrate into decision making on large-scale water resources projects.

**Key Issues and Challenges**

Deep learning captures the non-linear complex representations with datasets, making it a viable and efficient option in predictive model creation. In machine learning literature capturing the representation is known as representation learning (Goodfellow et al., 2016). Representation learning relies on artificial neural networks' ability in acting as a universal approximator (Hornik et al., 1989; Cybenko, 1989; Leshno et al., 1993) meaning ANNs with only one hidden layer, theoretically, could represent any function. The drawback regarding this ability is that the task in hand might need a hidden layer that is too large to still be feasible to be trained and executed. Adding new hidden layers to neural networks comes into play in order to cover this negation. Consequently, one important principle of representation learning is that, with minimal data engineering, the datasets should be fed to the neural network and let the neural network decide which features within the dataset are important towards the goal of representing that dataset. Even though this should be the case, the literature in the water domain does not widely apply this principle. We attribute this to the fact that datasets are neither extensive enough in terms of the number of given data, nor have vast spatial and/or temporal coverage.

As opposed to fields like computer vision and natural language processing, the water field lacks high quality, collected, curated, labeled, and published datasets that are used for benchmarking and method development. We identify this lack of benchmarking datasets as a key challenge (Ebert-Uphoff et al., 2017) which slows the state-of-the-science deep learning applications in water domain. Most studies reviewed herein acquire datasets from governmental agencies depending on their needs. Though potentially convenient for a contained effort of a single work, each one-off dataset dampens the speed of improvement of the state of the science in this field. If researchers had the opportunity to build on a widely-accepted dataset in their fields, they would be able to improve the accuracy of their models by taking advantage of previous models created using the same dataset. This collaboration around common datasets and models would open opportunities in the field, such as paving the way for their real-time usage. However, as we don't have many benchmarking datasets, many research groups all around the globe run similar networks on custom datasets they acquired with limited scientific interaction.

One other problem with the data provided by authorities is that they are dispersed among different agencies and they occasionally have mismatches in temporal and/or spatial coverage. Even though the data provided extends to many years before covering decades, one might need to access various databases created by several different government agencies in order to build a dataset. Further, the period of record commonly differs across agencies and areas which causes the data acquisition process time-consuming and sometimes inconclusive. Further complicating data acquisition, water data are suppressed due to their military significance in some strategic instances and contributes to limit progress in the field. We support growing access to governmental and agency-collected water data and its use in deep learning.

We consider the fundamental understanding of deep learning within the literature as another problematic issue emerging from the reviews made in this study. A common mistake regards what deep learning is and what deep learning is not. Most papers seem to interpret deep learning as a specific technique. However, deep learning is simply a broad term for various machine learning algorithms centered upon ANNs. Various studies reviewed here claim that they employ deep learning, yet they only take advantage of traditional ANN approaches. Most of the time this utilization doesn't surpass the extent of a study that employs conventional statistical modelling. This phenomenon raises questions whether these studies attempt to exploit the keywords, using the term "deep learning" to take advantage of the current scientific zeitgeist. We observe that these motives rarely result in work that forwards the deep learning literature. We particularly note the poor practice of ascribing work “deep learning” while not employing representation learning principles.

This comprehensive review identified that literature at the intersection of deep learning and water do not cover the used methods in detail. Most of the studies appear reluctant to give model/architecture details that are vital to reproduce the proposed training pipeline. Combined with the aforementioned dataset problems cause consistent barriers to reproducibility of work. This overshadows the reported accuracy of studies and slows advancement in the deep learning powered water field.

In contrast, some studies discuss unnecessary details over and over again in their manuscript. Discussing how the most efficient number of hidden layers or hyper-parameters are found by trial and error for many paragraphs without a hint of intuition, or discussing the optimal batch size for many paragraphs do not appear as the authors use pertinent deep learning theory. Not sharing intuition regarding technical choices also prevents the papers to reach their goal. There are many papers utilizing LSTM networks, but a few discuss the rationale behind their decision. One would at least expect a paper to describe why they choose LSTM networks over GRU networks in their setting. While this instance needs a comparison of two similar RNN implementations, it should also be noted that networks like GANs and DQNs find a very limited place for themselves in the literature. Furthermore, researchers use networks like ELM, ENN, and NAR in various tasks without sharing their relative success to more complex architectures widely-used in state of the art deep learning models.

**Ethics in DL Applications**

Broadly, ethics are concerned with describing the appropriateness of an act; whether something is “right” or “wrong” to do. Meanwhile, DL is a tool which can be used to produce algorithms which automate prediction *and* decision-making. Thus, most ethical concerns of DL derive from the central question: What will the DL application *do*? The ethical considerations of DL in water are no exception. Take, for example, a DL application that predicts streamflow within an urban catchment. By appearances, the act of streamflow prediction alone lacks any ethical character. However, if the same tool is then used to make decisions in disaster mitigation (Demir et al., 2018) or public planning, many questions with ethical dimensions arise (Ewing and Demir, 2020). What data were, and *were not*, used to develop the model? What biases exist in the dataset? How do these biases affect decision-making and human lives, and do the decisions reveal any discriminating behaviors? From this example, it is clear that the primary ethical considerations for the application of DL in water should be concerned with how entities -- people, the environment, communities -- will be affected by DL in decision-making workflows.

Though powerful, these DL tools simultaneously expand the reach and speed of decision-making, while also stripping away layers of context that would possibly be relevant to a humans’ decision-making process. However, many of the ethical decisions themselves in the water field remain unchanged and are primarily distributional; who receives water services, and their level of quality, and what level, and for whom, of risk is acceptable. These persistent ethical water dilemmas must be resolved within the new paradigm of DL in water.

As revealed by this literature review, few DL applications in the water literature include decision-making components. One paper, Rohmat et al. (2019), reviewed in this paper explicitly states that their DL tool is integrated into a decision support system. This lack of attention to DL in service of decision-making in water academia presents an opportunity for a new line of research. It is also an opportunity to incorporate the work of other fields early, such as the well-documented ethical concerns stemming from DL decision- and recommendation-engines in social and civil applications (such as policing, criminal justice, and self-driving cars (Angwin, Larson, Mattu, & Kirchner, 2016; O'Neil, 2016).) Further, a proliferation of AI Ethics frameworks have been developed, reviewed here (Hagendorff, 2019; Jobin et al., 2019), as well as numerous guidelines for applying and assessing algorithms in social and civil domains (O'Reilly, 2013; Rahwan, 2017; Reisman, Schultz, Crawford, & Whittaker, 2018). These guidelines stress a deep understanding of the task the algorithm is in service of, feedback on algorithm performance, and rigorous assessment procedures.

The discussions of ethics in the context of DL are part of a continued conversation of ethics generally. Questions of how to treat those in our communities, how to treat the environment, what is the "right" priority, or priorities, are not questions exclusive to the AI/DL domain. Rather, the new scale and speed provided by DL require the inspection of age-old ethical questions in a new light.

**Recommendations and Conclusions**

This paper provides a comprehensive review of the recent application of deep neural networks as novel solutions to tackle hydrological tasks and challenges. A total of 129 publications were systematically selected for rigorous review and analysis as grouped by their application area. Based on the statistical meta-analysis of journal publications dating between January 2018 to March 2020, it was empirically observed that the average number of deep learning applications per month in the water sector steadily increased in an exponential fashion. Further, the rapidly increasing body of work from these publications show deep learning's potential in a diversity of water sector applications. Key issues and challenges that may constitute setbacks and hindrances of deep learning adoption in the water industry have been identified and reported accompanied by recommendations to persevere in spite of the logistical, computational, expertise-related, and data-related challenges to its principled adoption. Based on the extent of this review and the broad spectrum of application areas, we anticipate the water sector will continue to incorporate deep learning at an accelerating rate and deep learning will play a key role in the future of water. Deep learning-powered technologies opened up a plethora of application and research opportunities to revolutionize hydrological science and workflow. In a bird's eye view fashion, key areas of innovation for future research include:

1. *Automated Forecasting*: As this review outlined, the majority of current deep learning applications in the hydrological domain focus on forecasting of numerous parameters (e.g. water level, discharge) given the problem's suitability for machine learning. In the future, efforts can be coordinated between agencies and research organizations to collaboratively develop complementary models that will yield actionable and reliable information. These models can be maintained and powered by a stream of real-time data influx to constitute the future of decision-making systems and geographical information systems.
2. *Published Datasets:* Lack of deep learning ready datasets within the water field was stressed in the previous section. The main problem caused by this absence of many datasets is that the research community does not build upon previous work in terms of constructing better neural network architectures and moving the state of art to the next iteration. This inference is supported by the fact that among the 30 papers related to flooding reviewed in this study there are only a few that use a previously curated, labeled dataset. The result is many papers are published that achieve the same task with almost identical methods but different data. This absence implicitly causes redundancy in the field. We believe if more studies focus on creating benchmark datasets that are open to researchers, both cumulativeness of the science would be satisfied and deep learning powered modeling in water resources research would go further in terms of generic applicability.

3. *AI as a Service:* As the popularity and usefulness of artificial intelligence tools increase, a new research area came to prominence in the computer science field. This area is focused on developing generalized and centralized web frameworks that can readily provide the means to develop custom AI solutions through Platform-as-a-Service (PaaS) systems. These systems hold great potential for the hydrological community as they allow developers to focus on designing and managing intelligent applications without the burden of infrastructure maintenance and adjustment of computing resources. They can provide intuitive graphical user interfaces to allow the development of hydrological deep learning applications by connecting predefined and custom models with provided datasets.
4. *Edge Computing:* The main propeller in the creation of smart applications is the consistent and diverse data flux. However, as the frequency and types of data resources expand, a centralized approach to collect and analyze data simply may not be viable for multivariate tasks. Furthermore, the costs associated with the transfer of large data from distributed sensors are not trivial and may discourage stakeholders to increase sensor coverage and data reporting interval. As a solution, edge computing offers a new perspective to process the data on the sensor. There is extensive research on utilizing deep learning for the internet of things through edge computing, which can allow the stakeholders of the water domain to innovate novel applications with existing or low-cost sensor networks. As a tangible example, a camera-equipped river monitoring sensor can employ deep learning to analyze pictures on the edge to detect any foreign objects (e.g. tree, human) on the river, and transmit only the useful information to a centralized system.
5. *Intelligent Assistants:* The massive amount of environmental and hydrological data makes it challenging to efficiently and effectively extract required knowledge in a timely manner. Manual efforts are often needed to analyze comprehensive raw data for decision-making purposes. As a solution, intelligent assistants serve as voice-enabled knowledge engines that can interpret a natural language question, apply human-like reasoning, and extract the desired factual response out of curated data. These assistants can be integrated into various communication channels for easy access including web systems, messaging applications (e.g. Skype, Messenger, SMS), personal assistants (e.g. Google Assistant, Apple Siri), home automation devices (e.g. Google Home, Amazon Alexa), augmented and virtual reality systems (e.g. HoloLens, Magic Leap, Oculus Quest), and automated workflow systems. The voice-enabled communication and immediate access to knowledge can facilitate hydrological research as well as natural disaster preparedness and response (Sermet and Demir, 2018a).
6. *Virtual and Augmented Reality*: Incorporation of deep learning with virtual and augmented reality environments provides a prominent research area due to its immersive nature that allows effective analysis of complex environmental phenomena that is not feasible to be orchestrated in real-life. For example, deep learning can power realistic flood simulations while mimicking human behavior to train first-responders, aid decision-makers, and educate the public (Sermet and Demir, 2018b). Another use case may be guiding on-site personnel that may require expertise, such as sensor maintenance, structural renovation, and field experiments, through heads-up displays and deep learning-powered recognition and decision support applications (Sermet and Demir, 2020).